\documentclass[
 reprint,
 amsmath,amssymb,
 aps,
 nofootinbib
]{revtex4-2}

\usepackage{graphicx}
\usepackage{dcolumn}
\usepackage{bm}
\usepackage{xcolor}
\usepackage{hyperref}
\usepackage[normalem]{ulem}
\usepackage{xtab,afterpage,longtable}

\hypersetup{
colorlinks=true,
citecolor=blue,
linkcolor=blue,
urlcolor=blue}

\def\apj{{ ApJ}}

\def\aap{{ A\&A}}

\def\mnras{{ MNRAS}}

\def\nat {{ Nature}}

\def\prd{{ Phys. Rev. D}}

\begin{document}

\preprint{APS/123-QED}

\title{Survival of ultraheavy nuclei in astrophysical sources: \\applications to protomagnetar outflows}

\author{Nick Ekanger$^{1,2,3}$}
\email{ekangernj@astr.tohoku.ac.jp}
\author{Mukul Bhattacharya$^{4,5,6}$}
\author{Kohta Murase$^{6,7,8,9}$}
\author{Shunsaku Horiuchi$^{3,10,11}$}

\affiliation{${}^1$Frontier Research Institute for Interdisciplinary Sciences, Tohoku University, Sendai 980-8578, Japan}
\affiliation{${}^2$Astronomical Institute, Graduate School of Science, Tohoku University, Sendai 980-8578, Japan}
\affiliation{${}^3$Center for Neutrino Physics, Department of Physics, Virginia Tech, Blacksburg, VA 24061, USA}
\affiliation{${}^4$Department of Physics, Wisconsin IceCube Particle Astrophysics Center, University of Wisconsin, Madison, WI 53703, USA}
\affiliation{${}^5$Department of Astronomy, Astrophysics and Space Engineering, Indian Institute of Technology Indore, Simrol, MP 453552, India}
\affiliation{${}^6$Department of Physics, The Pennsylvania State University, University Park, PA 16802, USA}
\affiliation{${}^7$Department of Astronomy \& Astrophysics, The Pennsylvania State University, University Park, PA 16802, USA}
\affiliation{${}^8$Institute for Gravitation and the Cosmos, The Pennsylvania State University, University Park, PA 16802, USA}
\affiliation{${}^9$Center for Gravitational Physics and Quantum Information, Yukawa Institute for Theoretical Physics, Kyoto University, Kyoto, Kyoto 606-8502, Japan}
\affiliation{${}^{10}$Department of Physics, Institute of Science Tokyo,
2-12-1 Ookayama, Meguro-ku, Tokyo 152-8551, Japan}
\affiliation{${}^{11}$Kavli IPMU (WPI), UTIAS, The University of Tokyo, Kashiwa, Chiba 277-8583, Japan}

\date{\today}

\begin{abstract}
Outflows of rapidly rotating protomagnetars have been considered as attractive sites for the synthesis of nuclei heavier than iron, but the question remains whether these nuclei are able to survive against photodisintegration as they make their way out of their formation environments. In this work, we present new analytic fitting formulae for the photodisintegration cross sections applicable to heavy nuclei beyond iron. We confirm that the results from the TALYS simulations are consistent with the theory of the giant dipole resonance, and apply the obtained new formulae to investigate whether ultraheavy nuclei entrained in protomagnetar outflows can be disintegrated by thermal and nonthermal photons before leaving the stellar envelope. We explore two outflow models: a spherical wind model and a jetted outflow model. For nuclei accelerated to the bulk speed of these outflows, their survival depends on the model and engine properties. For spherical winds, nuclei may survive for the first $\sim100\,{\rm s}$ post-core collapse, but as the wind Lorentz factor increases, the photodisintegration optical depth sharply rises and nuclei may no longer survive. For the jetted outflows arising from progenitors surrounded with stellar envelopes, nuclei can only survive before the jet breakout time in cases where the central engine has high spin-down energy, that is, with a high magnetic field strength and shorter spin period. In progenitors with more extended envelopes, the jet break out time is much longer, allowing for nonthermal photons to readily photodisintegrate nuclei in high spin-down energy cases. These results also outline some of the necessary, but not yet sufficient, conditions to source ultrahigh -energy cosmic-ray nuclei.
\end{abstract}

\maketitle

\section{Introduction}\label{sec:intro}

Ultrahigh energy cosmic rays (UHECRs), although observed for decades, are not fully understood (see Refs.~\cite{Blumer2009,KO2011,AB2019,Luis2019} for reviews). Most recently, their flux and energy spectrum have been observed through the Pierre Auger Observatory (PAO; e.g., \cite{PAO2015}) and Telescope Array (TA; e.g., \cite{AZ2013}). The Greisen-Zatsepin-Kuzmin (GZK; \cite{Greisen:1966jv,1966JETPL...4...78Z}) cutoff arises from the interaction of ultrahigh-energy protons with the cosmic microwave background (CMB), but the origin of the observed cutoff~\cite{Abraham2008,Abbasi2008} remains inconclusive. 

While the sources of UHECRs still remain unknown, recent observations have provided important clues. For instance, at the highest energies, UHECRs are more likely to be of intermediate-to-heavy nuclei (see, e.g., \cite{PierreAuger:2022atd}) and even ultraheavy nuclei may be present~\cite{Anchordoqui:1999aj,Zhang:2024sjp}, although the data cannot distinguish individual nuclei that are present. Nevertheless, what has become clear is that the data disfavors the UHECR to be entirely composed of protons or protons and helium nuclei (see, e.g., \cite{Jiang2021,KT2021}). Understanding their composition will in turn facilitate a better understanding of the astrophysical sources that can produce UHECRs. 

Proposed candidates of UHECR sources include jets from active galactic nuclei (AGNs, \cite{Norman1995,Dermer2009,Peer:2009vnw,TH2011,Murase:2011cy}), gamma-ray bursts including low-luminosity (LL) objects \cite{Waxman1995,MU1995,Murase2006,Murase:2008mr,Liu:2012zzd}, newly born rapidly rotating magnetars arising from core-collapse supernovae (CCSNe, \cite{Arons2003,KM2009,Fang:2012rx,Fang:2013vla}), and binary neutron star mergers~\cite{Takami:2013rza,Rodrigues:2018bjg,Murase:2018utn,Farrar:2024zsm,Zhang:2024sjp}. 
An important requirement for any UHECR source is that the
composition at the highest energies be enriched in nuclei relative to solar abundances \cite{Anchordoqui:1999aj}. In AGNs, such an enhancement may be achieved if jets reaccelerate pre-existing galactic cosmic rays, although this scenario can require tuning and remains uncertain~\cite{Caprioli:2015zka,Kimura:2017ubz,Mbarek:2019glq,Mbarek:2022nat}. For compact transients such as GRBs and engine-driven supernovae, as well as tidal disruption events (TDEs), it is natural that nuclei are supplied directly by their progenitor stars. If the progenitor is rapidly rotating, the injected nuclear composition shaped by stellar evolution can reproduce the observed UHECR spectrum and composition~\cite{Zhang:2017moz,Boncioli:2018lrv,Zhang:2018agl} (see Ref.~\cite{Zhang:2017hom} for TDEs involving white dwarfs). Moreover, nuclei are not only inherited from stellar nucleosynthesis but can also be explosively synthesized during the first few seconds after core bounce. In this context, protomagnetar outflows are especially attractive because they can synthesize substantial abundances of intermediate-mass and heavy nuclei \cite{Metzger2011b,Horiuchi2012,MBh2021,Ekanger:2022tia,Ekanger:2023mde}. 

Such relativistic outflows launched by rapidly rotating, highly magnetized protoneutron stars (PNSs) formed in massive-star collapse are also a leading scenario for long-duration GRBs \cite{Thompson2004}, making it plausible that the same engine both synthesizes heavy nuclei and accelerates them to ultrahigh energies. The resulting composition depends on the outflow properties (e.g., magnetization and time-dependent luminosity), which in turn follow from the spindown evolution of the central PNS (see \cite{MBh2021}).

In addition to providing and accelerating nuclei, a viable UHECR source must allow them to survive, i.e., escape without disintegration~\cite{Murase:2008mr,Wang:2007xj}. The main processes that can prevent heavy nuclei from escaping relativistic PNS outflows include photodisintegration, photomeson production, fragmentation, and nuclear spallation. Photodisintegration is particularly important because it shapes the maximum energy and the mass composition of nuclei during propagation through radiation fields \cite{1976puget,Stecker:1998ib,Khan:2004nd}. These considerations generally favor low-luminosity jets and/or magnetically dominated outflows~\cite{Murase:2008mr,Wang:2007xj,Metzger2011b,Horiuchi2012,Zhang:2017moz,Boncioli:2018lrv,Zhang:2018agl}. Low-power jets, including LL GRBs, may arise when relativistic jets are choked or partially smothered inside the progenitor envelope. The resulting phenomenology depends sensitively on the progenitor structure, e.g., Wolf--Rayet (WR) stars, blue supergiants (BSGs), and red supergiants (RSGs) (see Ref.~\cite{Bhattacharya:2022btx} and references therein). Such events can exhibit properties intermediate between classical GRBs and transrelativistic supernovae \cite{Soderberg2006}, offering a unified picture of the GRB--SN connection, e.g., Refs.~\cite{Margutti2013,Margutti2014}. 

Nucleus survival also affects multimessenger signatures: efficient photodisintegration can supply neutrons relevant for GeV neutrino production \cite{Murase:2013mpa,Carpio:2023wyt}, while stringent survival requirements can imply low ambient photon densities, thereby limiting high-energy neutrino production via photomeson interactions \cite{MuraseBeacom2010} or the escape of very high-energy gamma rays is guaranteed \cite{Murase:2008mr,Murase:2010va,Zhang:2023ewt}.

Motivated by this broader context, in this work we investigate whether nuclei synthesized via the $r$-process in protomagnetar-driven outflows can survive photodisintegration. Importantly, we focus on the fate of nuclei that are moving together with the bulk speed of the outflows, i.e., without assuming nonthermal components of ultraheavy nuclei, including UHECRs.   

The paper is organized as follows. In Sec.~\ref{sec:pd}, we examine photodisintegration cross sections in the dominant giant dipole resonance (GDR) channel and provide a new approximation applicable to heavy nuclei with $A\geq4$. In Sec.~\ref{sec:outflows}, we introduce two protomagnetar outflow models and identify the timescales on which their photon fields transition from thermal to nonthermal. In Sec.~\ref{sec:pdefficiency}, we compute the resulting photodisintegration optical depth and determine the conditions under which the nuclei can escape intact. We summarize the implications and discuss the remaining uncertainties in Sec.~\ref{sec:discussion} and Sec.~\ref{sec:summary}. 

\section{New Fitting Formulae of Photodisintegration Cross Sections for Heavy Nuclei}\label{sec:pd}

\begin{figure*}
\centering
\includegraphics[width=0.49\linewidth]{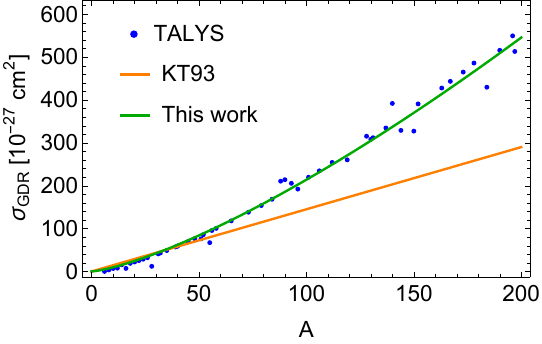}
\includegraphics[width=0.49\linewidth]{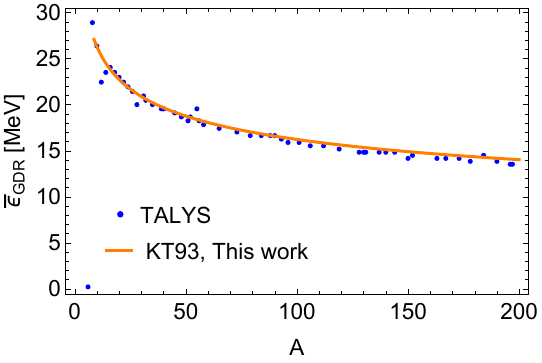}
\includegraphics[width=0.49\linewidth]{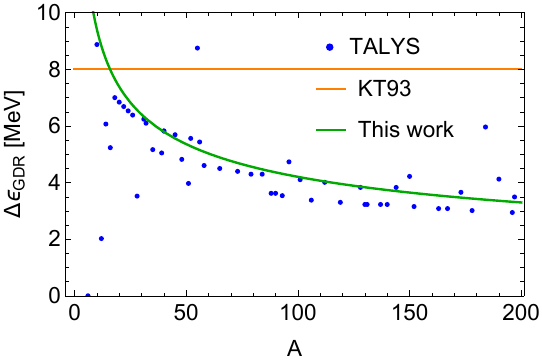}
\includegraphics[width=0.49\linewidth]{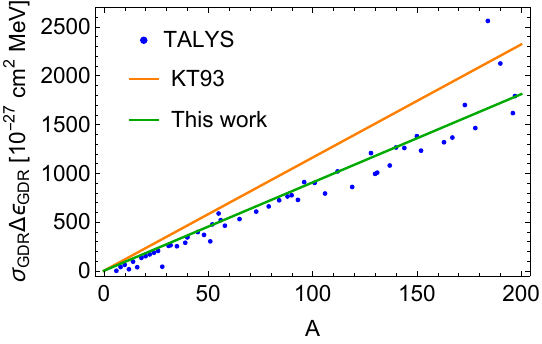}
\caption{ 
The TALYS data (shown as blue dots in each panel) compared to old fits (orange lines, see~\cite{KARAKULA1993229}~(henceforth KT93)) and our new fits (green lines in each panel, except for the \textit{top right} panel where our fit is the same as KT93). \textit{Top left:} the TALYS peak cross-section data ($\sigma_{\rm GDR}$, see Eq.~(\ref{GDRxs})) for GDR resonance. The best fit can be described by a power law $\sigma_{\rm GDR}\approx0.43A^{1.35}\times10^{-27}\,{\rm cm^2}$. \textit{Top right}: the best fit to TALYS data for the central energy ($\overline{\varepsilon}_{\rm GDR}$) of the GDR resonance, in which case the prior fit describes the data well. The best fit is given by $\overline{\varepsilon}_{\rm GDR}\approx42.65A^{-0.21}\,{\rm MeV}$. \textit{Bottom left}: the full-width at half-maximum ($\Delta\varepsilon_{\rm GDR}$) of the TALYS data versus $A$. The best fit is given by $\Delta\varepsilon_{\rm GDR}\approx21.05A^{-0.35}$, in contrast to a constant value of $8\,{\rm MeV}$ used in the previous studies. \textit{Bottom right}: the product $\sigma_{\rm GDR}\Delta\varepsilon_{\rm GDR}\approx9A\times10^{-27}\,{\rm cm^2~MeV}\propto\sigma_{A\gamma}$, resulting in an approximately linear $A$-dependence.}
\label{fig:xsupdateplot}
\end{figure*}

To assess whether nuclei formed through explosive nucleosynthesis can avoid disintegration in environments with a significant number density of high-energy photons, we first describe how the photodisintegration cross sections depend on the nuclei mass number and energy. Here, we introduce an updated parametrization for the GDR cross section. GDR is the largest (in terms of cross section) and typically the most important resonance among photodisintegration processes, although there are additional photodisintegration processes that are relevant at higher energies.

We use the TALYS simulation tool (version 1.95, see \cite{2012NDS...113.2841K,2019NDS...155....1K}) to calculate photodisintegration cross sections over the GDR and quasi-deuteron (QD) energy ranges (roughly $\sim1-100\,{\rm MeV}$). Previous studies \cite{Murase:2008mr,MuraseBeacom2010} approximated the GDR cross section as $\sigma_{A\gamma}\approx\sigma_{\rm GDR}\delta(\overline{\varepsilon}-\overline{\varepsilon}_{\rm GDR})\Delta\varepsilon_{\rm GDR},$ where $\sigma_{\rm GDR}$ is the peak value of the GDR approximation, $\overline{\varepsilon}_{\rm GDR}\sim42.65A^{-0.21}\,{\rm MeV}$ is the energy at which the resonance occurs and $A$ is the nuclei mass number, and $\Delta\varepsilon_{\rm GDR}$ is the resonance width. Previous studies like Ref.~\cite{KARAKULA1993229} (KT93) base this approximation on nuclear data up to iron ($Z=26$, $A=56$). To improve on this existing approximation by analyzing data from nuclei heavier than iron as well, we use TALYS to simulate the photonuclear cross sections for nuclei up to gold ($Z=79$, $A=197$). We use this TALYS data to then determine the approximate height of the GDR resonance ($\sigma_{\rm GDR}$), peak energy for resonance ($\overline{\varepsilon}_{\rm GDR}$), and its width ($\Delta\varepsilon_{\rm GDR}$), where this width is defined as the full width at half maximum of the peak. This is a valid approximation since the majority of the GDR cross section outputs given by TALYS are roughly symmetric-Lorentzian functions.

We find that $\overline{\varepsilon}_{\rm GDR}\approx 42.65A^{-0.21}\,{\rm MeV}$ fits the TALYS data for the central energy of the GDR resonance well, as shown in the \textit{top right panel} of Fig.~\ref{fig:xsupdateplot}. In the same figure, we also show new fits to $\sigma_{\rm GDR}$ (in the \textit{top left panel}), $\Delta\varepsilon_{\rm GDR}$ (in the \textit{bottom left panel}), and the product of the two since this is proportional to $\sigma_{\rm A\gamma}$ (in the \textit{bottom right panel}). From the variation of $\sigma_{\rm GDR}$ with mass number $A$, we find that $\sigma_{\rm GDR}$ scales with an approximately linear dependence on the mass number up to iron ($A=56$). However, a power law with $\sigma_{\rm GDR} \propto A^{4/3}$ is a better fit for larger mass numbers up to $A\sim200$. For example, in Ref.~\cite{Murase:2008mr}, a constant $\Delta\varepsilon_{\rm GDR}\approx8\,{\rm MeV}$ is assumed, but here we update this to a different power law given by $\approx 21.05A^{-1/3}\,{\rm MeV}$. 

The best-fit parametrizations are as follows: $\sigma_{\rm GDR}\approx 0.43\times10^{-27}A^{1.35}\,{\rm cm^2}$, $\overline{\varepsilon}_{\rm GDR}\approx42.65A^{-0.21}\,{\rm MeV}$, $\Delta\varepsilon_{\rm GDR}\approx21.05A^{-0.35}\,{\rm MeV}$, such that the GDR cross section scales linearly with mass number,
\begin{align}\label{GDRxs}
    \nonumber \sigma_{A\gamma}&\approx\sigma_{\rm GDR}\Delta\varepsilon_{\rm GDR}\delta(\overline{\varepsilon}-\overline{\varepsilon}_{\rm GDR})\\
    \nonumber &\approx(0.43A^{1.35}\times10^{-27}\,{\rm cm^2})\delta(\overline{\varepsilon}-42.65A^{-0.21}\,{\rm MeV})\\
    &\times(21.05A^{-0.35}\,{\rm MeV}).
\end{align}
These best-fit scaling parameters are also listed in Table~\ref{tab:xsresults}, where we compare our results with previous approximations. Note that in our results, the total cross section is linear with mass number $A$, but the prefactor and width have different $A$ dependencies. Ref.~\cite{Firestone:2020rty} finds $\sigma_{\rm GDR}\sim0.48A^{4/3}\times10^{-27}\,{\rm cm^2}$ which is very similar to the value of $0.43A^{1.35}\times10^{-27}\,{\rm cm^2}$ obtained in this work, although different values of $\overline{\varepsilon}_{\rm GDR}$ and $\Delta\varepsilon_{\rm GDR}$ are inferred.

\renewcommand{\arraystretch}{1.5}
\begin{table*}
    \centering
    \begin{tabular}{|c|c|c|c|c|c|}
    \hline
    &$\mathbf{\sigma_x}$&\textbf{Prefactor} $\mathbf{[10^{-27}~{\rm \bf{cm^2}}]}$&\textbf{Center [MeV]}&\textbf{Width [MeV]}&$\mathbf{A}$\textbf{--range}\\
    \hline
    \emph{This work}&$\sigma_{\rm GDR}$&$0.43A^{1.35}$&$42.65A^{-0.21}$&$21.05A^{-0.35}$&$4\leq A\leq197$\\ \hline
    \emph{\citet{KARAKULA1993229}}&$\sigma_{\rm GDR}$&$1.45A$&$42.65A^{-0.21}$&8.0&$56\geq A>4$\\
         &$\sigma_{\rm GDR}$&$1.45A$&$0.925A^{-2.433}$&8.0&$A\leq4$\\ \hline
    \end{tabular}
    \caption{Comparison of best-fit scaling parameters for cross section approximations relevant in this work for the disintegration of nuclei. Here, the GDR approximation is of the form $\sigma_x\approx{\rm (Prefactor)}\delta(\overline{\varepsilon}-{\rm Center})$(Width). $A$-range describes the range of mass numbers for which the approximation is valid.}
    \label{tab:xsresults}
\end{table*}

%KM added
In the Goldhaber-Teller picture \cite{goldhaberteller1948}, GDR is the dominant isovector collective mode in which the proton and neutron fluids oscillate out of phase as two slightly displaced liquid drops, producing a large electric dipole (E1) moment. It is known that the E1 strength is largely concentrated in this collective resonance. The contribution from the GDR region to the energy-weighted E1 sum is a large fraction of the total Thomas--Reiche--Kuhn (TRK) sum rule. Since the TRK sum scales as $NZ/A$, where $N$ is the number of neutrons, the integrated photodisintegration cross section is $\int \sigma_{A\gamma}\,d\bar{\varepsilon} \propto NZ/A$. For heavy nuclei, this gives $NZ/A\sim A/4$ and hence $\int \sigma_{A\gamma}\,d\bar{\varepsilon}\approx \sigma_{\rm GDR}\Delta\bar{\varepsilon}_{\rm GDR} \propto A$. If the relevant restoring force is governed mainly by bulk nuclear properties, the corresponding frequency is $\Gamma\propto R^{-1} \propto A^{-1/3}$, leading to $\overline{\varepsilon}_{\rm GDR}\propto A^{-1/3}$. If the dominant damping is controlled by surface-related decoherence, one expects $\Delta\bar{\varepsilon}_{\rm GDR}=\Gamma\sim \hbar/\tau \propto R^{-1} \propto A^{-1/3}$, and therefore $\sigma_{\rm GDR}\propto A^{4/3}$ is expected. 

Although the best-fit approximation for $\sigma_{\rm GDR}$ and $\Delta\varepsilon_{\rm GDR}$ agree well with the TALYS data points, there are some notable outliers - especially in the case of $\Delta\varepsilon_{\rm GDR}$ for smaller mass numbers. Some nuclei have cross sections that are not well described by Lorentzian functions and/or have double peaks, such as carbon-14, oxygen-16, silicon-28, vanadium-51, and manganese-55. Most nuclei, however, are well described by a single Lorentzian function.

We do not consider higher-energy photodisintegration resonances, photomeson production and fragmentation, because GDR is the predominant photonuclear interaction in the photon energy range of interest. Further, TALYS does not simulate resonances beyond the pion production threshold. Tools like the Geant4 toolkit~\cite{GEANT4:2002zbu} and other analytical formulae can model these processes, but are not considered here (see \cite{1996PhDT........59R,Allard:2005ha,Murase:2008mr} for the impacts of higher-energy photodisintegration processes). Spallation is another process that can lead to the destruction of nuclei and may have a minor impact on the results of nucleus survival for some scenarios (see Sec.~\ref{sec:discussion} for discussion on this point).

\begin{figure*}
\centering
\includegraphics[width=\linewidth]{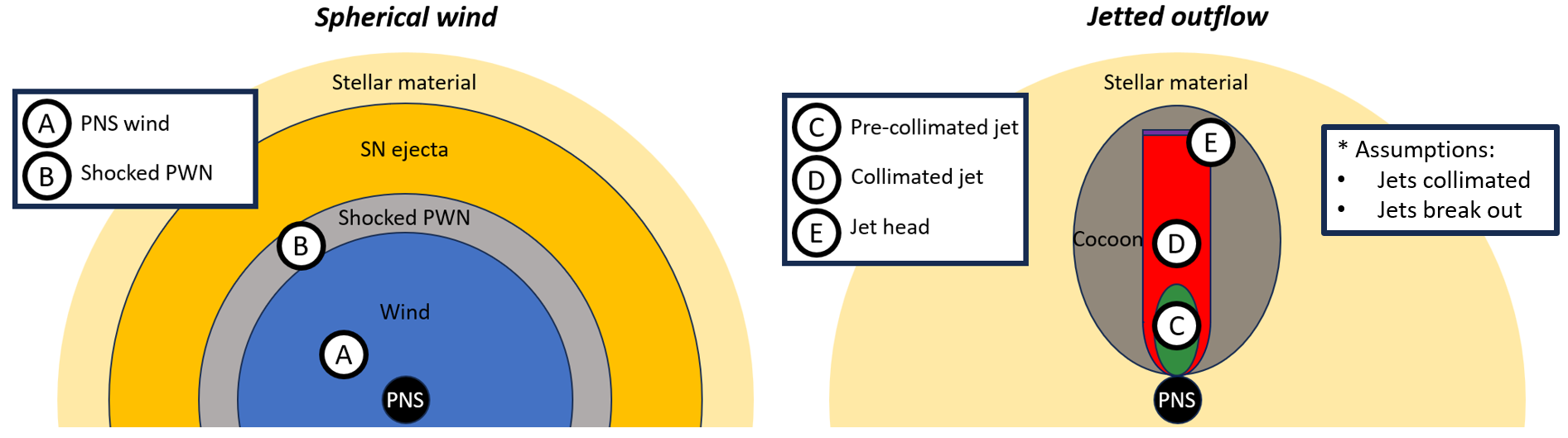}
\caption{Schematic picture of the two models we consider. \textit{Left panel:} a spherical wind originating from the protomagnetar central engine. In Region A, a neutrino-driven wind flows from the PNS central engine, which eventually catches up to the nonrelativistic pulsar wind nebula in Region B. \textit{Right panel:} a jetted outflow which may also originate from protomagnetar central engines. In Region C, a pre-collimated jet expands into Region D, which is a collimated jet region. Region E is the jet head, which may break out of the surrounding stellar material.}
\label{fig:schematic}
\end{figure*}

\section{Applications to protoneutron star outflows}\label{sec:outflows}

The cross section formulae we provide can be used for many astrophysical sources. In this section, we explore two models of protomagnetar-driven outflows: the spherical wind and jetted outflow models. Figure \ref{fig:schematic} shows the schematic picture of these two models. Below, we determine the evolution of the (non)thermal nature of the photon spectra. Then, in Sec.~\ref{sec:pdefficiency}, we will discuss the time evolution of the photodisintegration optical depth.

\subsection{Photon analysis framework}\label{subsec:survival}

Within the wind, nuclei are exposed to high-energy photons which may disintegrate them into nucleons. In order to determine whether the nuclei get destroyed, we first determine the energy distribution of these photon fields. To do this, we calculate the Thomson optical depth, $\tau_T$, to determine whether photons are thermalized due to copious electron-positron pairs in the wind. 
This is given by
\begin{equation}\label{thomson}
\tau_T\approx y_{\pm}n_x\sigma_TR_x/\Gamma_x,
\end{equation}
where the subscript `x' describes either the spherical wind (w) or jetted outflow (h) model (see subsections below). Additionally, $y_{\pm}=(n_x+n_{\pm})/n_x$ is the pair production enhancement factor where $n_x=\Dot{M}_x/(4\pi R_x^2m_pc\Gamma_x)$ is the wind baryon number density and $n_{\pm}=\Dot{M}_{\pm}/(4\pi R_x^2m_ec\Gamma_x)$ is the pair number density, $\Gamma_x$ is the Lorentz factor of the outflow, $\Dot{M}_{\pm}\approx(2.5\times10^{-17}\,M_{\odot}~\textrm{s}^{-1})\mu_{\pm,6}B_{\rm dip,15}P_{i,-2}^{-2}R_{\rm NS,6}^3$ is the Goldreich-Julian (GJ) mass-loss rate \cite{1969ApJ...157..869G}. Here, $\mu_{\pm}\approx10^6$ is the pair multiplicity \cite{bucciantini11,Timokhin:2015dua}, $B_{\rm dip}$ is the dipole magnetic field strength and $P_i$ is the spin period of the PNS. The Thomson cross section is given by $\sigma_T=6.65\times10^{-25}\,{\rm cm^2}$. In this study, we consider outflows with $10^{13}\,{\rm G}<B_{\rm dip}<10^{15}\,{\rm G}$ and $1\,{\rm ms}<P_i<30\,{\rm ms}$.

The distribution of the photon spectrum will have quantitative impacts on nuclear survivability. To help understand the evolution of the photon spectrum, we label the approximate time when photons transition when the system is optically thin as $t_{\rm Th}$; this timescale is defined as the time when $\tau_T=1$. This should be understood as a proxy because the details depend on radiative transfer calculations. Nevertheless, it will suffice for the outflow models we consider. In the following sections, we explore a spherical wind model and a jetted outflow model that are driven by rapidly rotating, highly magnetized protoneutron stars. Whether particles can remain as undestroyed nuclei and escape the system depends, then, on: \textit{(A)} the timescale after which the outflow transitions from thermal to nonthermal photon-dominated system ($t_{\rm Th}$), and \textit{(B)} the timescale at which the nuclei are transported outside the progenitor. 

\subsection{Spherical wind}\label{subsec:windevolution}

The schematic picture of the spherical wind is shown in the \textit{left panel} of Fig.~\ref{fig:schematic} with two regions: the neutrino-driven wind (Region A) and the shocked pulsar wind nebula (PWN, Region B). In this picture, the relativistic wind (A) catches up to the nonrelativistic PWN (B) over time. In this section, we describe how we calculate $\tau_T$ (Eq.~\ref{thomson}) in both these regions.

In Region A, we calculate the wind radius $R_w$ and ejecta radius $R_{\rm ej}$ by numerically solving the differential equations \cite{Murase:2014bfa,Kashiyama2016},
\begin{eqnarray}
\frac{dR_w}{dt}&=&\sqrt{\frac{7}{6(3-\delta)}\frac{\mathcal{E}_{\rm tot}}{M_{\rm ej}}\left(\frac{R_w}{R_{\rm ej}}\right)^{3-\delta}}+\frac{R_w}{t}\\
\frac{dR_{\rm ej}}{dt}& \equiv &V_{\rm ej} = \sqrt{\frac{2\mathcal{E}_{\rm tot}}{M_{\rm ej}}},
\end{eqnarray}
where $\delta=1$ is used for the density profile of the ejecta. We assume an ejecta mass of $M_{\rm ej}=3M_\odot$ and SN explosion energy of $\mathcal{E}_{\rm tot}=10^{51}\,{\rm erg}$. We set $R_w(t=0) = R_{\rm ej}(t=0) = R_{\rm LC} = cP_i/2\pi$ as the initial conditions, where $R_{\rm LC}$ is the radius of the light cylinder. 

The mass-loss rate from the PNS surface due to the wind driven by neutrinos is given by \cite{Metzger2011a},
\begin{align}\label{Mdot}
    \nonumber \Dot{M}_w&= (5\times10^{-5}~M_\odot~\textrm{s}^{-1})\left[\frac{L_{\nu}}{10^{52}\,{\rm erg~s^{-1}}}\left(\frac{\varepsilon_{\nu}}{10\,{\rm MeV}}\right)^2\right]^{5/3}\\
    &\times \mathcal{F}_{\rm mag}\left(C_{\rm inel}\frac{R_{\rm NS}}{10^6\,{\rm cm}}\right)^{5/3}\left(\frac{M_{\rm NS}}{1.4\,M_{\odot}}\right)^{-2}.
\end{align}
Several correction factors are accounted for here, including $\mathcal{F}_{\rm mag}=f_{\rm op}f_{\rm cen}$ which considers the fraction of the PNS surface threaded by open magnetic field lines ($f_{\rm op}$) and an enhanced mass loss rate due to magnetorotational slinging ($f_{\rm cen}$), $C_{\rm inel}$ for inelastic neutrino-electron scatterings, and a stretch factor of $\eta_s=3$ to model longer PNS cooling timescales due to rapid rotation (for these, see Ref.~\cite{Metzger2011a}). $L_{\nu}$ is the $\nu_e+\bar{\nu}_e$ neutrino luminosity, $\varepsilon_{\nu}$ is the mean neutrino energy, $R_{\rm NS}$ is the PNS radius, and $M_{\rm NS}$ is the PNS mass. We choose $M_{\rm NS}=1.4\, M_{\odot}$ and $R_{\rm NS}=10^6\,{\rm cm}$ based on Ref.~\cite{Pons1999}. The PNS mass outflow rate eventually transitions from being baryon-dominated to pair-dominated. This occurs when $\Dot{M}_x=\Dot{M}_{\pm}$ is attained (hereafter, labeled with $t=t_{\rm GJ}$). Thus, if $t_{\rm Th}>t_{\rm GJ}$, the nuclei are likely to face thermal photons, while if $t_{\rm Th}<t_{\rm GJ}$, nuclei can also encounter nonthermal photons.

The Lorentz factor of the wind, $\Gamma(r=R_w)=\Gamma_w$, can be expressed as~\cite{Drenkhahn2002},
\begin{equation}
\label{LorentzFactor}
\Gamma=
\begin{cases} 
\sigma_0(r/R_{\textrm{mag}})^{1/3}, & r\leq R_{\textrm{mag}} \\
      \sigma_0, & r>R_{\textrm{mag}} 
\end{cases},
\end{equation}
where $r$ is evaluated at $R_w$ and $R_{\rm mag}$ is the magnetic dissipation radius given by $R_{\rm mag}\approx(5\times10^{12}\,{\rm cm})(\sigma_0/10^2)^2(P/{\rm ms})(\epsilon/0.01)^{-1}$. We parametrize the reconnection velocity with $\epsilon \sim 0.01$~\cite{Metzger2011a}. Wind magnetization is $\sigma_0=\phi_B^2\Omega^2/\Dot{M}_wc^3$, where $\phi_B=(f_{\rm op}/4\pi)B_{\rm dip}R_{\rm NS}^2$ is the magnetic flux from a rotating dipole field (of strength $B_{\rm dip}$) and $\Omega$ is the PNS angular velocity.

\begin{figure*}
\centering
\includegraphics[width=0.49\linewidth]{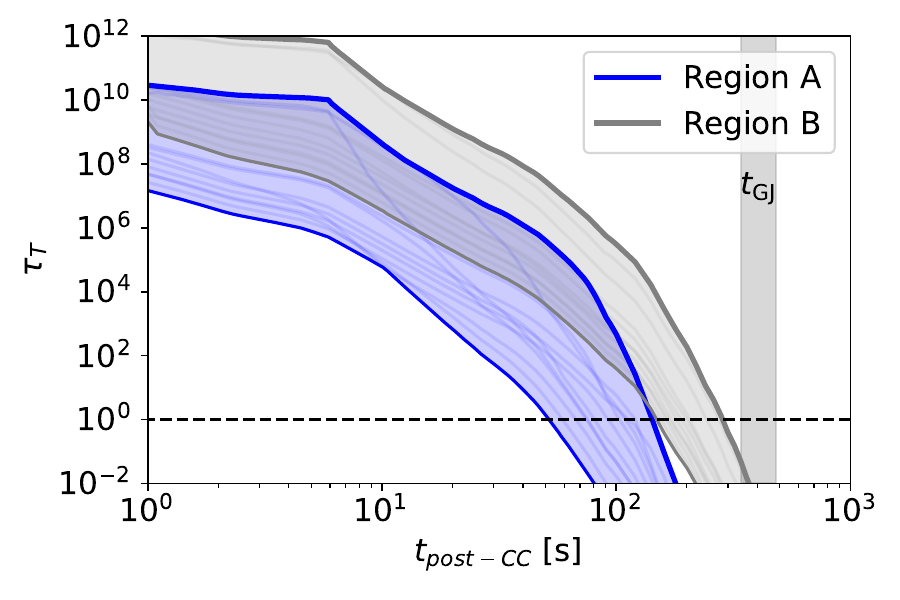}
\includegraphics[width=0.49\linewidth]{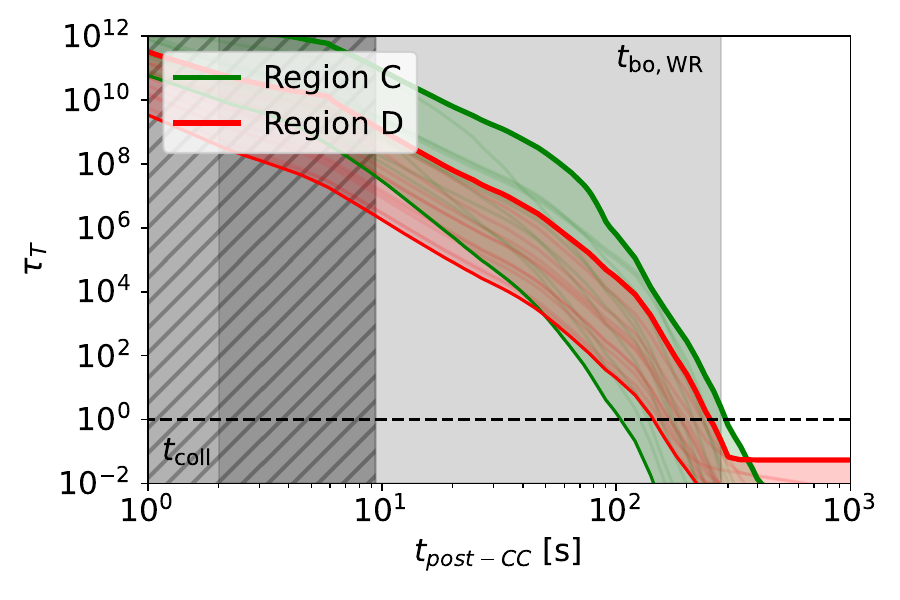}
\caption{Thomson optical depth, $\tau_T$, for several cases is shown as a function of time. The dashed black line shows $\tau_T=1$, which we consider the transition point at which photon fields are thermal ($\tau_T>1$) or nonthermal ($\tau_T<1$). The shaded regions represent the minimum to maximum $\tau_T$ values that are calculated for our range of $B_{\rm dip}$ and $P_i$ configurations and the translucent lines between are the individual results for each configuration. The gray hatched region represents the time when the jets become collimated ($t_{\rm coll}$) and the gray shaded region is the breakout time for WRs ($t_{\rm bo,WR}$). For all cases, $t_{\rm coll}<t_{\rm bo}$. \textit{Left panel}: $\tau_T$, for the spherical wind case for Regions A (blue shaded) and B (gray shaded). The vertical gray shaded region shows the range in $t_{\rm GJ}$ values for our range of $B_{\rm dip}$ and $P_i$ configurations. For all configurations of $B_{\rm dip}$ and $P_i$, $t_{\rm Th}<t_{\rm GJ}$, implying that nonthermal photons are expected by $t_{\rm GJ}$. \textit{Right panel}: $\tau_T$ for the jetted outflow case of a WR progenitor in Regions C (green shaded) and D (red shaded). After $t_{\rm bo,WR}$, nuclei can freely escape the star. For WRs, $t_{\rm Th}>t_{\rm bo,WR}$ (except for one case), for BSGs, $t_{\rm Th}>t_{\rm bo,BSG}$ for rapidly rotating engines, and for all RSG cases, $t_{\rm Th}<t_{\rm bo,RSG}$.}
\label{fig:tautcomparison}
\end{figure*}

In Region B, we also calculate the Thomson optical depth at $R_x=R_w$, assuming $\dot{M}_x=\dot{M}_w$, but because the outflow is nonrelativistic, $\Gamma_x=1$ and $n_x=\dot{M}_w/(4\pi R_w^2m_pV_w)$ where $V_w=dR_w/dt$.

The \textit{left panel} of Fig.~\ref{fig:tautcomparison} shows $\tau_T$ as a function of time post core-collapse for the spherical wind model in the shaded blue region (Region A) and in the shaded gray region (Region B). The width of this shaded region reflects the values for the $B_{\rm dip}$ and $P_i$ configurations used in this work, where the edges stand for the minimum and maximum values of $\tau_T$. The horizontal dashed black line denotes $\tau_T=1$ when the photon distribution transitions from thermal to nonthermal, which generally occurs around $100-300\, {\rm s}$. Finally, the shaded gray region shows the range in times $t=t_{\rm GJ}\sim300-500\,{\rm s}$, corresponding to the range in $B_{\rm dip}$ and $P_i$, at which the outflow mass becomes dominated by electron-positron pairs instead of baryons. For all spherical wind models, we find $t_{\rm Th}<t_{\rm GJ}$, such that nuclei in the outflow can encounter nonthermal photons before the mass outflow rate drops considerably to be pair dominated.

\subsection{Jetted outflow}\label{subsec:jetted outflow}

If the outflow propagates through dense stellar material, it may become collimated. A cocoon forms as the outflow interacts with the surrounding medium and the subsequent magnetized outflow can burrow through the star like a jet. Such different outflow geometries can lead to different distributions of photons. Such a scenario is shown in the \textit{right panel} of Fig.~\ref{fig:schematic}. Here, region C is the pre-collimated jet outflow, Region D is the collimated jet, and Region E is the jet head. In this work, we do not consider the survival of nuclei in the cocoon region or the jet head, but only in the pre-collimated and collimated jets. Typically, the cocoon's energy density is low enough for nucleus survival~\cite{Horiuchi2012}. Because the outflows are jetted, the number density of the baryons is increased by a factor of $(\theta_j^2/2)^{-1}$ \cite{Bhattacharya:2022btx}. We choose a jet with a fixed opening angle of $\theta_j=10^{\circ}$ at the jet launch site near the PNS.

In Region C, $R_x=R_{\rm cs}$, or the radius of the collimation shock (see Sec.~3.3 in Ref.~\cite{Bhattacharya:2022btx}), $\Gamma_x=\Gamma_j=\Gamma(r=R_{\rm cs})$, and $n_x=\dot{M}_w/(4\pi R_{\rm cs}^2m_pc\Gamma_j)/(\theta_j^2/2)$. In the collimated jet of Region D, $R_x=\min(R_*,~R_h)=R_h$ (until the jet breaks out of the progenitor), $\Gamma_x=\Gamma_{\rm cj}=1/\theta_j$, and $n_x=\dot{M}_w/(4\pi R_h^2m_pc\Gamma_{\rm cj})/(\theta_j^2/2)$. The cocoon that forms can be approximately modeled as a cylinder, whose head radius is described as $R_h=c\xi_h\beta_h t$, where $\xi_h=(5-\alpha)/3$ is a coefficient, $\alpha\sim2-3$ is the power-law index of the density profile, and $\beta_h$ is the velocity of the jet head (see \cite{Bhattacharya:2022btx}, Eq.~(7)). As the figure shows, we have implicitly assumed collimation. While jets begin as uncollimated, we find that all become collimated (when $R_h>R_{\rm cs}$) quickly within $t_{\rm coll}<10\,{\rm s}$. The exact timing depends on both outflow properties as well as the stellar progenitor. For the former, we consider outflows with $10^{13}\,{\rm G}<B_{\rm dip}<10^{15}\,{\rm G}$ and $1\,{\rm ms}<P_i<30\,{\rm ms}$. For the latter, we consider a smaller/denser WR star and an extended/sparser blue and red supergiants (BSG, RSG). 

We estimate the jet breakout time, $t_{\rm bo}$, for WR progenitors using the analytical expression from the text after Eq.~(1) in \cite{MBh2021} (see also, Eq.~(12) in \cite{Ekanger:2022tia}). This expression takes into account the $B_{\rm dip}-P_i$ configuration and signifies when the magnetized outflow has met the minimum energy requirement necessary to break out of the stellar surface (see also \cite{Bromberg:2014sja}). Our WR, BSG, and RSG progenitor stars have radii of $R_*=5.15\,R_{\odot}$, $52\,R_{\odot}$, and $875\,R_{\odot}$, respectively (see \cite{Bhattacharya:2022btx}, for the details of their density profiles and how they lead to different jet structures). As a simple estimate, we calculate the breakout times for BSGs and RSGs using the ratio of their radii to WR radii, i.e., $t_{\rm bo,BSG}\sim10\times t_{\rm bo,WR}$ and $\sim170\times t_{\rm bo,WR}$, because the velocity of the jet head does not change significantly over time. Jets may fail to break out of the star, if they do not meet the energy requirements to do so. For WR stars, this tends to occur for central engines with smaller magnetic fields (for jets with $\theta_j=20^{\circ}$, this occurs for $B_{\rm dip}\lesssim3\times10^{15}\,{\rm G}$, roughly independent of $P_i$ \cite{Bhattacharya:2022btx}). For BSGs and RSGs, this further depends on the stellar density structure and, for these reasons, we do not consider choked jets in this work.
% In this study, we do not consider choked jets that do not breakout of their WR progenitors; these jets may arise from central engines that do not have sufficient isotropic jet luminosity to break out ($B_{\rm dip}\lesssim3\times10^{15}\,{\rm G}$, roughly independent of $P_i$) \cite{Bhattacharya:2022btx}. 
% Note on choked jets here: this depends on energy argument, which goes as thetaj^4, so reducing thetaj from 20deg to 10deg here increases ratio by 16 and may help our argument. unclear if RSG/BSG situation is better or worse

The jet breakout time has an impact on the window over which the nuclei are exposed to energetic photons within the jet. If $t_{\rm Th}>t_{\rm bo}$, the photons are thermalized even after breakout. Since the jet breaks out of the star before the surrounding photons become nonthermal, the nuclei only interact with lower energy thermal photons. On the other hand, if $t_{\rm Th}<t_{\rm bo}$, the photons can contain a nonthermal component before breakout. Timescales, as in the spherical wind case, are dependent on $B_{\rm dip}$ and $P_i$, but also have a strong dependence on the progenitor's stellar radius.

Figure~\ref{fig:tautcomparison} (\textit{right panel}) shows $\tau_T$ as a function of time post core-collapse for the jetted outflow models of WR progenitors. In green we show the range of optical depth values for Region C, and in red we show the range of values for Region D. The dark gray hatched region at early times is the time over which the jets become collimated and, in this region, our calculations are not valid. The gray shaded region shows the range of breakout times $t_{\rm bo,WR}$ for various $B_{\rm dip}-P_i$ configurations. To check the validity of our calculations up to breakout time, we confirm that $t_{\rm coll}<t_{\rm bo}$ for all jets.

For all outflows in WRs (except the $B_{\rm dip}=10^{13}\,{\rm G}$, $P_i=30\,{\rm ms}$ case), $t_{\rm Th}>t_{\rm bo,WR}$. This suggests that nuclei only encounter thermal photons before breaking out and escaping the stellar envelope. For BSGs, $t_{\rm Th}>t_{\rm bo,BSG}$ for the rapidly rotating cases we consider. Thus, nuclei in the slowly rotating progenitors encounter nonthermal photons before the jet breaks out. For RSGs, $t_{\rm Th}<t_{\rm bo,RSG}$ for all cases, and the photons can be nonthermal in all jets that we consider. Further, $t_{\rm Th}<t_{\rm GJ}$, where $t_{\rm GJ}\sim300-500\,{\rm s}$ as in the \textit{left panel} of Fig.~\ref{fig:tautcomparison}. The high $B_{\rm dip}$ configurations tend to transition to nonthermal photons earlier, primarily due to a lower number density. Thus, progenitors with smaller envelopes do not typically undergo the transition from thermal to nonthermal photons prior to breakout. The survival consequences because of this are explored in Sec.~\ref{sec:pdefficiency}.

\section{Effective photodisintegration optical depth}\label{sec:pdefficiency}

\begin{figure*}
\centering
\includegraphics[width=0.49\linewidth]{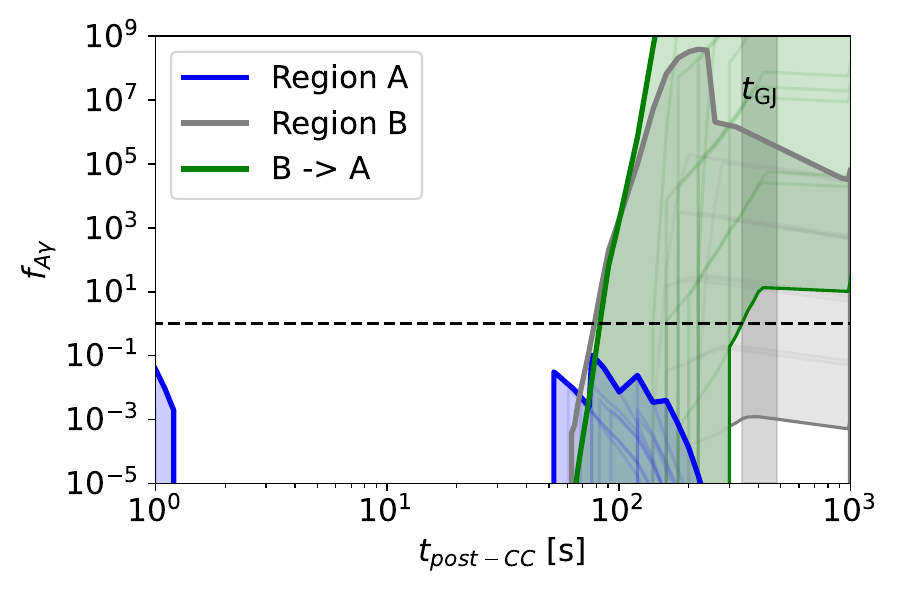}
\includegraphics[width=0.49\linewidth]{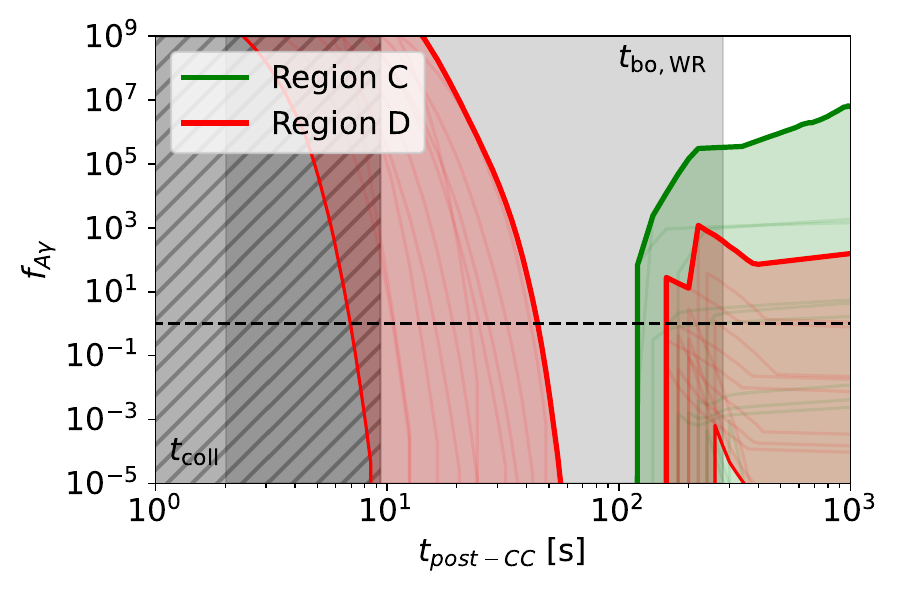}
\includegraphics[width=0.49\linewidth]{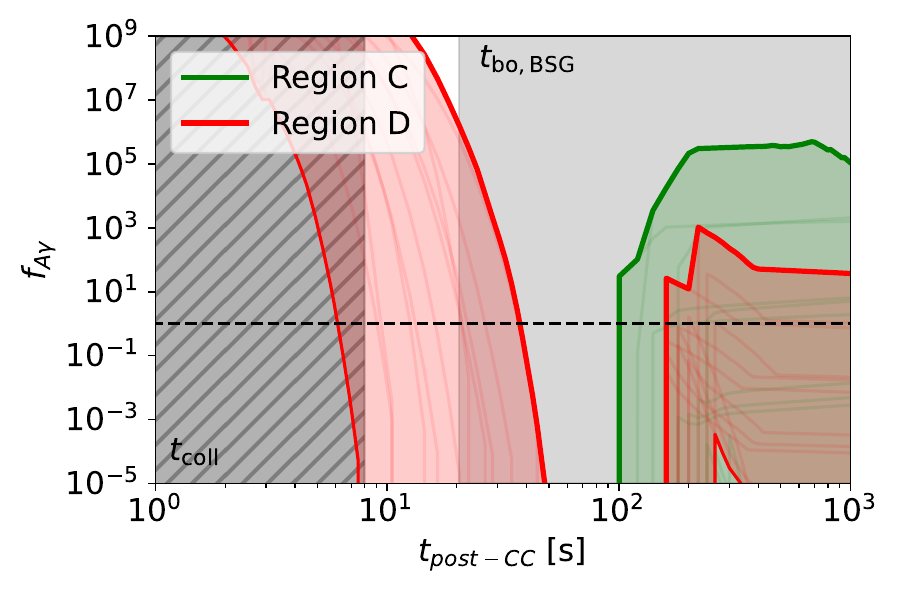}
\includegraphics[width=0.49\linewidth]{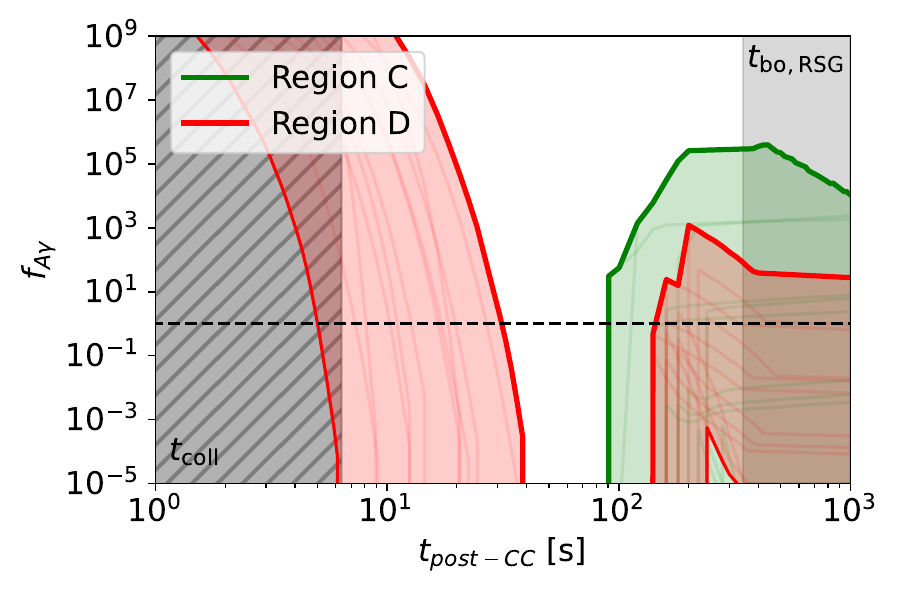}
\caption{The photodisintegration optical depth, $f_{A\gamma}$, is shown as a function of time post core-collapse. In each panel, the value of $f_{A\gamma}=1$ (which we consider as the boundary of ``survival'' vs.~``destruction'') is denoted by the horizontal dashed black line assuming a pure-iron composition ($A=56$). The nuclei energy is assumed to be $\Gamma_xm_Ac^2$, where $\Gamma_x$ is the bulk Lorentz factor in each Region. The shaded colored regions show the extent of values when varying the PNS magnetic field and spin period, whose boundaries are given by the maxima and minima values found. The lines within these regions correspond to the individual $B_{\rm dip}/P_i$ configurations. The hatched $t_{\rm coll}$ region shows the time when the jets become collimated, before which the calculations are unphysical. Finally, the shaded gray regions show the extent of breakout times for each case. \textit{Top left panel:} $f_{A\gamma}$ for the spherical wind case. For the first $\sim100\,{\rm s}$ ($t_{\rm Th}$ for Region A), $f_{A\gamma}\ll1$ so nuclei can survive in both regions. Around this time, photons become nonthermal, but the energy density is not high enough in Region A to cause photodisintegration. However, very quickly, due to leaked photons in Region A and Region B, nuclei are efficiently destroyed. \textit{Top right panel:} $f_{A\gamma}$ for the jetted outflow case from WR progenitors. In some high $P_i$ cases, $f_{A\gamma}\gg1$ in the regime where thermal photons can destroy nuclei before the breakout time. In all other cases, nuclei can survive photodisintegration before breaking out of the star. \textit{Bottom left panel:} $f_{A\gamma}$ for the jetted outflow case from BSG progenitors. In this scenario, $f_{A\gamma}<1$ for all cases at $t_{\rm bo,BSG}$ (except the case of $B_{\rm dip}=10^{15}\,{\rm G}$, $P_i=10\,{\rm ms}$, where $f_{A\gamma}\sim1$). \textit{Bottom right panel:} $f_{A\gamma}$ for the jetted outflow case from RSG progenitors. In this case, the breakout time is much longer. In this nonthermal photon regime, nuclei in high spin-down energy cases are readily photodisintegrated.}
\label{fig:efficiencyplot}
\end{figure*}

In Sec.~\ref{sec:outflows}, we evaluated the timescale  ($t_{\rm Th}$) at which different outflow models transition from a thermal photon field to potentially hosting also a nonthermal photon field and the time at which the nuclei may escape their environment. In this section, we investigate nucleus survival and the extent to which this depends on the photon distribution. To do this, we calculate the photodisintegration optical depth as a function of time.

The effective photodisintegration optical depth is estimated as
\begin{equation}
    f_{A\gamma}=t_{A\gamma}^{-1}t_{\rm dyn},
\end{equation}
where $t_{\rm dyn}=R_x/(\Gamma_xc)$ is the dynamical timescale of the outflow and the photodisintegration timescale is (e.g., Ref.~\cite{Murase:2008mr})
\begin{equation}\label{timescale}
    t_{A\gamma}^{-1}=\frac{c}{2\gamma_A^2}\int_{\overline{\varepsilon}_{\rm th}}^\infty d\overline{\varepsilon}\ \sigma_{A\gamma}(\overline{\varepsilon})\kappa_{A\gamma}(\overline{\varepsilon})\overline{\varepsilon}\int_{\overline{\varepsilon}/2\gamma_A}^\infty d\varepsilon\ \varepsilon^{-2}\frac{dn}{d\varepsilon}.
\end{equation}
Here, $\gamma_A=E_A/m_Ac^2$ is the nucleus Lorentz factor, $\overline{\varepsilon}_{\rm th}\sim10\,{\rm MeV}$ is the photodisintegration threshold energy, $\sigma_{A\gamma}$ is the photodisintegration cross section, $\kappa_{A\gamma}\sim1/A$ is the nucleus inelasticity, $dn/d\varepsilon$ is the photon energy spectrum that the nuclei are exposed to, $\varepsilon$ is the photon energy in the comoving frame and $\overline{\varepsilon}$ denotes the photon energy in the nucleus rest frame. In this work, we assume that nuclei are accelerated to the bulk Lorentz factor of each region, i.e., $E_A=\Gamma_xm_Ac^2$.

Prior to $t=t_{\rm Th}$, $\tau_T$ is much larger than unity, resulting in a thermal photon distribution described by,
\begin{equation}\label{thermaldist}
    \frac{dn}{d\varepsilon}=\frac{8\pi}{(hc)^3}\frac{\varepsilon^2}{e^{\varepsilon/k_BT_{\gamma}}-1},
\end{equation}
where $T_{\gamma}$ is the photon temperature, related to the gamma-ray energy density by $U_\gamma=aT_\gamma^4$. This distribution results in an analytical expression for $f_{A\gamma}$,
\begin{equation}\label{efficiencytherm}
f_{A\gamma}=\frac{4\pi c\sigma_{\rm GDR}\Delta\varepsilon_{\rm GDR}\overline{\varepsilon}_{\rm GDR}}{A\gamma_A^2(hc)^3}k_BT{\rm ln}{(1-e^{-y})}^{-1}t_{\rm dyn},
\end{equation}
where $y=\overline{\varepsilon}_{\rm GDR}/(2\gamma_Ak_BT)$. We suggest that nuclei undergo significant disintegration when $f_{A\gamma}\gtrsim1$.

After $t_{\rm Th}$, $\tau_T<1$ and we describe the nonthermal distribution with a broken power law:
\begin{equation}\label{nonthermaldist}
    \frac{dn}{d\varepsilon}=\mathcal{N}\frac{(1-e^{-\tau_{\gamma\gamma}})}{\tau_{\gamma\gamma}}
    \begin{cases}
        (\varepsilon/\varepsilon_{\rm break})^{-\alpha},&\varepsilon_{\rm min}<\varepsilon<\varepsilon_{\rm break}\\
        (\varepsilon/\varepsilon_{\rm break})^{-\beta},&\varepsilon_{\rm break}\leq\varepsilon\\
    \end{cases}
\end{equation}
where $\mathcal{N}$ is the normalization, $\varepsilon_{\rm min}\sim\varepsilon_{\rm ssa}\sim1\,{\rm eV}$ is the minimum photon energy allowed by synchrotron self-absorption \cite{Murase:2008mr}, $\varepsilon_{\rm break}$ is the break energy, and $\tau_{\gamma\gamma}$ is the optical depth of $e^{+}e^{-}$ pairs created through photon annihilation \cite{Murase:2022vqf}. We choose $\varepsilon_{\rm break}=10^{2.5}\,{\rm keV}$ in the comoving frame and the low/high photon spectra indices of $\alpha=1$ and $\beta=2.2$ \cite{Murase:2008mr}.  The normalization of the photon spectrum, $\mathcal{N}$, is chosen such that the radiation energy density, $U_{\gamma}=\int \varepsilon~dn/d\varepsilon~d\varepsilon$, is always conserved. To finally calculate the photodisintegration optical depth, we need to define the energy density of the photons and dynamical timescale in each region.

\subsection{Spherical wind}
In Region A, the energy density is related to the baryonic mass-loss of the neutrino-driven wind: $U_{\gamma,A}=\epsilon_{\gamma}\Gamma_w(\dot{M}_w+\dot{M}_{\pm})c^2/(4\pi \Gamma_w^2R_w^2c)$, where $\epsilon_\gamma=0.3$ is the fraction of the energy converted to radiation. The dynamical timescale is given by $t_{\rm exp,A}=R_w/(\Gamma_wc)$. In Region B, however, the energy density is proportional to the spin-down luminosity of the protoneutron star engine: $U_{\gamma,B}=\epsilon_eL_{\rm spin}/(4\pi R_w^2c)$, where $\epsilon_e\sim1$ is the fraction of energy converted to electrons and positrons, and $L_{\rm spin}$ is the electromagnetic spin-down luminosity (or equation 5) of Ref.~\cite{Murase:2014bfa}. The dynamical timescale is given by $t_{\rm exp,B}=R_w/V_w$. Finally, photons leak from Region B into Region A, with an energy density of $U_{\rm \gamma,BA}=f_{\rm BA}U_{\rm \gamma,B}\Gamma_w^2$, where $f_{\rm BA}=\min(1/\tau_{T,\rm B},~1)$ is the leakage fraction.

In Fig.~\ref{fig:efficiencyplot}, we show $f_{A\gamma}$ as a function of time post-core collapse for our two physical model scenarios. In each panel, the dashed black line denotes $f_{A\gamma}=1$, a nucleus mass of $A=56$ is assumed, and vertical gray shaded regions are marked for other relevant timescales. The lines inside the shaded regions show the results for each configuration considered in our $B_{\rm dip}$ and $P_i$ space, while the boundaries of the shaded regions show the maxima and minima results across all cases. In the \textit{top left panel}, we show the results for the spherical wind model. In Region A, $f_{A\gamma}<1$ from the internal energy of the photons in that region. Although nuclei in Region A escape disintegration for $\lesssim100\,{\rm s}$, the photons leaking into the region are able to disintegrate nuclei efficiently after this point. Around $t_{\rm GJ}$, nuclei are efficiently disintegrated, especially for high $B_{\rm dip}$ and low $P_i$ protoneutron stars, but $f_{A\gamma}\sim10$ around this time for the $B_{\rm dip}=10^{13}\,{\rm G}$ and $P_i=30\,{\rm ms}$ case. In Region B, similarly, $f_{A\gamma}>1$ only after $\gtrsim100\,{\rm s}$, after which the nuclei can survive disintegration in the low $B_{\rm dip}$/high $P_i$ cases. These suggest that PNS with higher energy density - derived from their higher magnetorotational energy - are more efficient at disintegrating nuclei. The sharp rise in photodisintegration optical depth around $\sim100\,{\rm s}$ is because of the transition of photons from thermal to nonthermal spectra as well as a large increase in $\Gamma_w$ occurring around the same time.

\begin{figure*}
\centering
\includegraphics[width=0.49\linewidth]{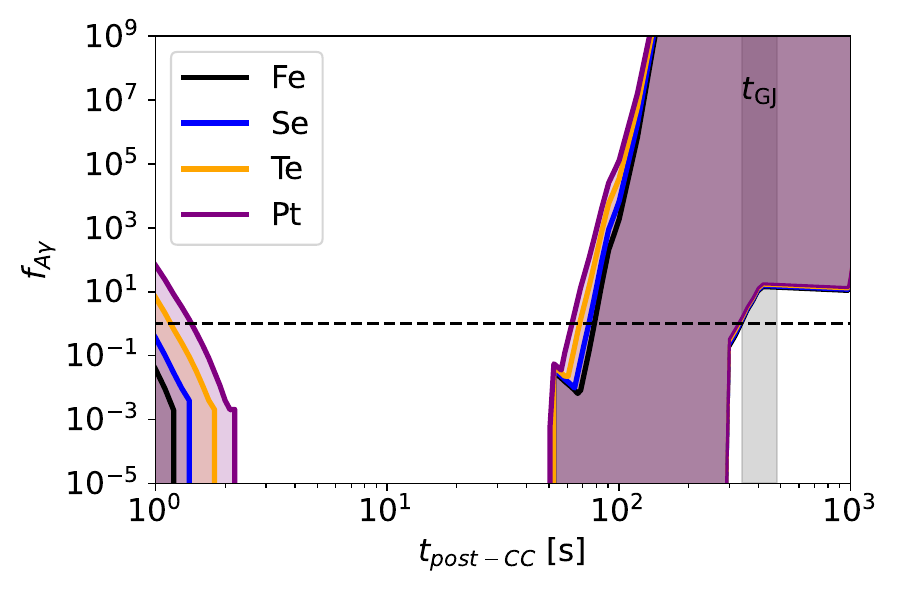}
\includegraphics[width=0.49\linewidth]{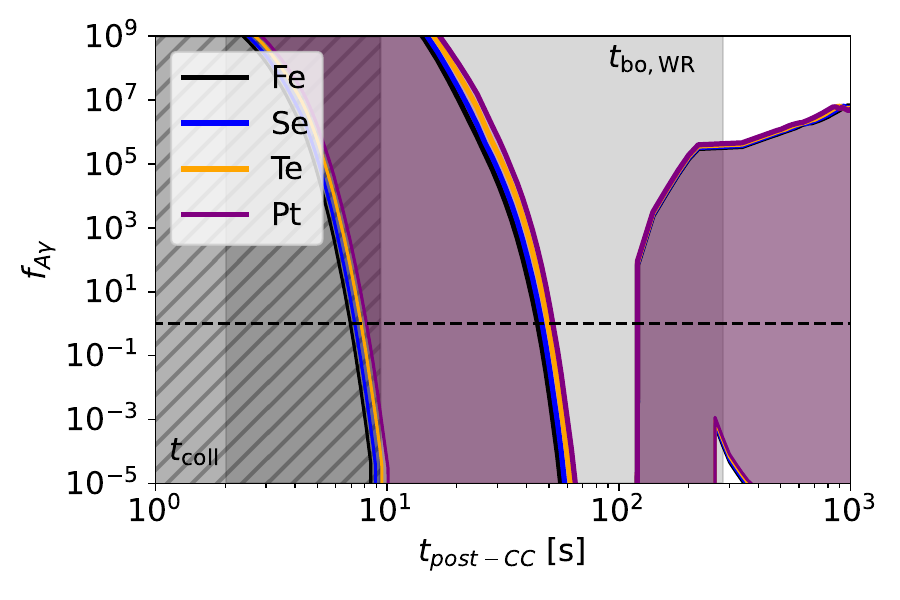}
\caption{Photodisintegration optical depths, where $f_{A\gamma}=1$ denotes our survival criterion, are shown as a function of time for our spherical wind model (\textit{left panel}) and jetted outflow model (WR case, \textit{right panel}), with varying nuclei composition. In black, we show the result for iron, which is the same result as the top row of Fig.~\ref{fig:efficiencyplot}, to compare with other cases. In blue, orange, and purple, we also show the efficiencies for selenium, tellurium, and platinum. This shows there is a slightly positive correlation of $f_{A\gamma}$ with mass number, and heavier nuclei are somewhat less likely to survive in general.}
\label{fig:efficiencycomp}
\end{figure*}

\subsection{Jetted outflow}

In Region C, we consider the photons that are leaked from Region D: $U_{\rm \gamma,C}=f_{\rm DC}U_{\rm\gamma,D}\Gamma_{\rm DC}^2$, where $f_{\rm DC}=\min(1/\tau_{T\rm,D},~1)$ is the leakage fraction from Region D to C and $\Gamma_{\rm DC}=\Gamma_{\rm j}\Gamma_{\rm cj}(1-\beta_{\rm j}\beta_{\rm cj})$ is the relative Lorentz factor between the pre-collimated jet and the collimated jet.  The dynamical timescale is given by $t_{\rm exp,C}=R_{\rm cs}/(\Gamma_j c)$. In Region D, $U_{\rm\gamma,D}=4\Gamma_{\rm DC}^2n_jm_pc^2$. This is valid for $\tau_{T\rm,D}\gg1$, and is realized for about $\sim200-300\,{\rm s}$, so evaluation of results for the WR cases is reasonable (since $t_{\rm bo,WR}\lesssim200-300\,{\rm s}$), but should be noted in some BSG cases and the RSG case. Because the jet is collimated, the dynamical timescale is given by the advection timescale: $t_{\rm adv,D}\sim R_h/(\Gamma_h c)$ (although, $\Gamma_h\sim1$).

In the other three panels, we plot $f_{A\gamma}$ for WR, BSG, and RSG progenitors in the \textit{top right}, \textit{bottom left}, and \textit{bottom right}, respectively. Although the values of $f_{A\gamma}$ do not change considerably depending on the progenitor, the breakout time, $t_{\rm bo}$, changes dramatically between these systems. This has implications for nucleus survival, as nuclei can escape freely after the breakout time. Prior to $t_{\rm bo,WR}$, $f_{A\gamma}\gg1$ in Region D (and negligible in Region C) for high $P_i$ cases, while $f_{A\gamma}$ decreases far below unity for low $P_i$ cases. In this regime, photons are still thermal. For BSG progenitors, there is only one case where $f_{A\gamma}>1$ before $t_{\rm bo,BSG}$: $B_{\rm dip}=10^{15}\,{\rm G}$ and $P_i=10\,{\rm ms}$. In this one case, $f_{A\gamma}$ is only slightly larger than unity and occurs after $t_{\rm Th}$, when photons are nonthermal. Although there are several cases in the figure where $f_{A\gamma}\gg1$ after $\sim100\,{\rm s}$ in Regions C and D, this occurs in the scenarios with high PNS spindown energy. In such scenarios, the jets breakout very quickly, so nuclei are able to escape before this time. In the RSG case, $t_{\rm bo,RSG}\gtrsim300\,{\rm s}$ is much later. Thus, nuclei can be exposed to nonthermal photons at late times and experience efficient photodisintegration. This disintegration occurs primarily for the high spindown energy cases (short $P_i$ and higher $B_{\rm dip}$). Note that these calculations are performed only up to $1000\,{\rm s}$, however the breakout times of RSGs could be longer than this.

There are some similarities in the results between the spherical wind and jetted outflow scenarios. In general, it seems that at early times, when photons are thermal, the high spindown energy cases (with high $B_{\rm dip}$ and low $P_i$) are more efficiently photodisintegrated. However, the results at later times is more nuanced and depends on the details of nuclei escape in each model.

Based on these results, the distribution of the ambient photon field can have major consequences on the survivability of nuclei. Previous works generally found that nonthermal photon distributions result in photodisintegration timescales that are limited by the dynamical/expansion timescales of magnetized outflows. The photon distribution is not well known, which would affect the outcome of nucleus survival. Synchrotron photons in the wind would be in the fast cooling regime \cite{Murase:2013mpa,Carpio:2023wyt}, while a broken power-law spectrum or the Band function can also be used in the jet \cite{Murase:2008mr,Giannios:2006jb,MBh2021}. The choice of minimum and maximum cutoff energies somewhat affects the energy at which $f_{A\gamma}$ decreases.

Because we determine new analytical approximations for the photodisintegration cross sections of heavy nuclei in Sec.~\ref{sec:pd}, we can estimate the efficiency of disintegration for heavy nuclei as well. This may be an important process to consider if outflows of supernovae produce heavy elements. In Fig.~\ref{fig:efficiencycomp}, we show how these results depend on mass number. In this figure, we only show the spherical wind model and WR jetted outflow models to show how efficiency scales with mass number. The results in black, for ``Fe'' or iron are the same as the results in Fig.~\ref{fig:efficiencyplot}. We also show, in blue, orange, and purple, the results for selenium (``Se'', $A=79$), tellurium (``Te'', $A=128$), and platinum (``Pt'', $A=197$). As expected, $f_{A\gamma}$ increases monotonically with mass number, due to the mass dependence in the GDR cross section and because of the increased energy of the nuclei in the outflows, since they have energies of $\Gamma_xm_Ac^2$.

\section{Discussion}\label{sec:discussion}

There have been several studies that quantify the survival of nuclei in outflows from compact objects. Ref.~\cite{Horiuchi2012} finds that the survival of thermal nuclei in outflows depends on the luminosity of the source. Generally, lower luminosity sources have lower photon energy densities and nuclei are more easily able to survive. The nucleus survival is regarded as a necessary condition for UHECRs to escape their environments. This has been studied for GRB producing systems \cite{Wang:2007xj,Murase:2008mr,Murase:2010va}. Ref.~\cite{MBh2021} focuses on magnetar-driven outflows \cite{MBh2021} and finds that high spin-down energy cases efficiently disintegrate nuclei at small radii. We find, however, that this trend is reversed at larger radii, where leaking photons in the optically thin regime efficiently disintegrate nuclei. Because of the modeling of multiple zones and acceleration to bulk Lorentz factors, not ultra-high energies, our results somewhat differ from these other studies.

Our study isolates the role of GDR photodisintegration for nuclei that are accelerated primarily by the bulk motion of the outflow, and it adopts a survival criterion ($f_{A\gamma}=1$). Several important physical effects were not treated self-consistently and should be addressed in future work.

We assumed that nuclei are co-moving with the outflow and thus have characteristic energy $\Gamma_x m_A c^2$. However, the nuclei could be accelerated to even higher energies via internal shocks, shear, or magnetic reconnections (see, e.g., Ref.~\cite{MBh2021}). Higher-energy nuclei interact with different parts of the photon spectrum and may enter regimes where additional channels (e.g., photomeson production) become dominant. These channels may become important and compete with GDR when nuclei are accelerated to relativistic energies and in cases where even higher spin-down energy magnetars drive outflows, where the bulk Lorentz factor is larger than what is considered here. In such cases, the window of survival will likely be further restricted, but still depends on the competition of timescales like jet breakout time. A self-consistent treatment coupling acceleration, radiation, and nuclear interactions is required to assess survival in such scenarios.

Our photodisintegration cross section mainly models the symmetric Lorentzian part of the GDR resonance, but in reality, the quasi-deuteron process somewhat increases the cross section at energies just higher than the GDR peak energy. Thus, our box approximation slightly underestimates the true photodisintegration optical depth at these energies by a factor of $\lesssim2$. Nuclei like oxygen, silicon, and some iron-group elements have additional resonances in the GDR region. When we fit our box approximation to these nuclei, our estimated width is relatively larger than single resonance nuclei, thus these atypical nuclei overcome the slight underestimate of the true optical depth that usual nuclei show. This effect does not affect our qualitative results. Another approximation like that of Ref.~\cite{kossov} yield similar results to our functional form that updates the commonly used and convenient analytical form. Note that the assumption of constant widths (see the \textit{lower right panel} of Fig.~\ref{fig:xsupdateplot}) is also sufficient to yield qualitative results but mass-dependent widths better reproduce the TALYS simulation results.

Using $f_{A\gamma}=1$ as a binary survival criterion neglects intermediate regimes in which only a fraction of nuclei may disintegrate. For $0.1\lesssim f_{A\gamma}\lesssim 1$, single interactions can create a mixed composition, while for $1\lesssim f_{A\gamma}\lesssim 10$ repeated interactions can drive nuclei toward intermediate masses without complete destruction. Tracking the time-dependent mass and charge distribution (rather than a survival/no-survival outcome) would be necessary to connect to the observable CR composition, especially if candidate sources preferentially produce moderately heavy nuclei. We focused on the GDR channel, but at very high or ultrahigh energies and/or in nonthermal photon fields, where accelerated nuclei will interact with considerably harder photon fields, photomeson production and fragmentation can also become important \cite{Murase:2008mr,Globus:2014fka,Boncioli:2016lkt,Morejon:2019pfu} and warrants additional analysis.

In such a parameter-driven study, there are several uncertainties, whose effect on results should be noted. In particular, the competition of timescales, like $t_{\rm bo}$, is important. This breakout time depends somewhat strongly on $B_{\rm dip}$ and $P_0$ \cite{MBh2021} and, in practice, the measurements of these quantities have larger than $\gtrsim10\%$ uncertainty \cite{Lu:2014oda}. In this sense, model uncertainty represents dominant systematics that could result in qualitative and quantitative differences. However, we maintain that our models are useful for illuminating central engine trends and are commonly discussed in the literature, e.g., Refs.~\cite{Metzger2011a,Murase:2013mpa,Lu:2014oda,MBh2021,Ekanger:2022tia,Bhattacharya:2022btx}. RSG stars are less impacted by this uncertainty, since $t_{\rm bo}$ is generally much larger than the dynamical and disintegration timescales. Spectral uncertainties, like the nonthermal peak energy and softness, have the potential to change the efficiency of disintegration due to higher-energy resonances, but our choices are supported by literature. Finally, uncertainties in parameters like pair multiplicity, reconnection speed, and GDR fitting formulae, are not expected to significantly change the main conclusions.

Spallation from nucleus--nucleus collisions can compete with photon-driven destruction when the baryon density is high and relative energies approach nuclear binding scales. Although we estimated that the spallation optical depth $\tau_{\rm sp}$ can be comparable to $f_{A\gamma}$ for our benchmark models, we did not model spallation self-consistently. A detailed composition study should include both photodisintegration and spallation, and should assess implications for $r$-process nucleosynthesis and for the emergence of intermediate-mass nuclei (see also Refs.~\cite{Horiuchi2012,David:2015ura,Morejon:2019pfu}).

If heavy nuclei are efficiently destroyed, the outflow becomes neutron rich and free neutrons may decouple dynamically from the charged component. Previous works have explored neutrino production from inelastic collisions of decoupled neutrons \cite{Murase:2013mpa,Carpio:2023wyt}, typically assuming proton--neutron-dominated composition. Our results suggest that this assumption may hold after $\sim 100\,{\rm s}$ for spherical winds but can be strongly parameter dependent in jets, especially given the interplay among $t_{\rm Th}$, the decoupling time, and $t_{\rm GJ}$. A time-dependent model that tracks composition together with the radiation field and dynamical evolution is needed to make robust multimessenger predictions.

\section{Summary}\label{sec:summary}

Photodisintegration is one of the central concerns for modeling the sources of heavy nuclei. Candidate sources include GRBs, AGNs and TDEs, which are also considered as potential sources of heavy-nuclei UHECRs. In this work, based on TALYS simulations, we presented the updated GDR cross-section fits that are applicable to heavy nuclei up to $A=197$. 

We applied our formulae to magnetized outflows from protomagnetars, where the nuclei may be subject to bulk acceleration, and quantified photodisintegration efficiencies in a time-dependent manner. Across a broad range of central-engine parameters and progenitor structures, we found that nucleus survival is controlled by the competition between the system's transition timescale (set by $t_{\rm GJ}$ in spherical winds and by $t_{\rm bo}$ in jets) and the epoch of photon thermalization (modeled by $t_{\rm Th}$), as well as by jet collimation effects that increase photon densities. These results delineate which combinations of engine parameters and progenitor envelopes can plausibly preserve heavy nuclei prior to their escape into the interstellar medium and constrain their contribution to the metal population within galaxies. They also provide a baseline for future, more complete treatments that couple particle acceleration, radiation, and full composition evolution. Furthermore, extending the present framework to different sources with their distinct radiation fields and dynamical timescales will be important for establishing whether sources of high-energy nuclei can preserve heavy or intermediate-mass composition up to the time of their escape.

\begin{acknowledgments}
We thank Hamid Hamidani, Michael Unger and B. Theodore Zhang for useful discussions.  
This work was supported by NSF Grants Nos. AST-1908689 (K.M. and M.B.), AST-2108466 (K.M.), AST-2108467 (K.M.), AST-2308021 (K.M.), and PHY-2209420 (S.H.). We also acknowledge KAKENHI grant Nos. 20H01901 (K.M.), 20H05852 (K.M.), and 23H04899 (S.H.). 
M.B. acknowledges support from the Eberly Research Fellowship at the Pennsylvania State University and the Simons Collaboration on Extreme Electrodynamics of Compact Sources (SCEECS) Postdoctoral Fellowship at the Wisconsin IceCube Particle Astrophysics Center (WIPAC), University of Wisconsin-Madison. 
This work was supported by World Premier International Research Center Initiative (WPI Initiative), MEXT, Japan.
\end{acknowledgments}

\section*{Data Availability}
The data are available from the authors upon reasonable request.

\bibliography{main}

\begin{thebibliography}{85}%
\makeatletter
\providecommand \@ifxundefined [1]{%
 \@ifx{#1\undefined}
}%
\providecommand \@ifnum [1]{%
 \ifnum #1\expandafter \@firstoftwo
 \else \expandafter \@secondoftwo
 \fi
}%
\providecommand \@ifx [1]{%
 \ifx #1\expandafter \@firstoftwo
 \else \expandafter \@secondoftwo
 \fi
}%
\providecommand \natexlab [1]{#1}%
\providecommand \enquote  [1]{``#1''}%
\providecommand \bibnamefont  [1]{#1}%
\providecommand \bibfnamefont [1]{#1}%
\providecommand \citenamefont [1]{#1}%
\providecommand \href@noop [0]{\@secondoftwo}%
\providecommand \href [0]{\begingroup \@sanitize@url \@href}%
\providecommand \@href[1]{\@@startlink{#1}\@@href}%
\providecommand \@@href[1]{\endgroup#1\@@endlink}%
\providecommand \@sanitize@url [0]{\catcode `\\12\catcode `\$12\catcode
  `\&12\catcode `\#12\catcode `\^12\catcode `\_12\catcode `\%12\relax}%
\providecommand \@@startlink[1]{}%
\providecommand \@@endlink[0]{}%
\providecommand \url  [0]{\begingroup\@sanitize@url \@url }%
\providecommand \@url [1]{\endgroup\@href {#1}{\urlprefix }}%
\providecommand \urlprefix  [0]{URL }%
\providecommand \Eprint [0]{\href }%
\providecommand \doibase [0]{https://doi.org/}%
\providecommand \selectlanguage [0]{\@gobble}%
\providecommand \bibinfo  [0]{\@secondoftwo}%
\providecommand \bibfield  [0]{\@secondoftwo}%
\providecommand \translation [1]{[#1]}%
\providecommand \BibitemOpen [0]{}%
\providecommand \bibitemStop [0]{}%
\providecommand \bibitemNoStop [0]{.\EOS\space}%
\providecommand \EOS [0]{\spacefactor3000\relax}%
\providecommand \BibitemShut  [1]{\csname bibitem#1\endcsname}%
\let\auto@bib@innerbib\@empty
%</preamble>
\bibitem [{\citenamefont {{Bl{\"u}mer}}\ \emph {et~al.}(2009)\citenamefont
  {{Bl{\"u}mer}}, \citenamefont {{Engel}},\ and\ \citenamefont
  {{H{\"o}randel}}}]{Blumer2009}%
  \BibitemOpen
  \bibfield  {author} {\bibinfo {author} {\bibfnamefont {J.}~\bibnamefont
  {{Bl{\"u}mer}}}, \bibinfo {author} {\bibfnamefont {R.}~\bibnamefont
  {{Engel}}},\ and\ \bibinfo {author} {\bibfnamefont {J.~R.}\ \bibnamefont
  {{H{\"o}randel}}},\ }\bibfield  {title} {\bibinfo {title} {Cosmic rays from
  the knee to the highest energies},\ }\href
  {https://doi.org/10.1016/j.ppnp.2009.05.002} {\bibfield  {journal} {\bibinfo
  {journal} {Progress in Particle and Nuclear Physics}\ }\textbf {\bibinfo
  {volume} {63}},\ \bibinfo {pages} {293} (\bibinfo {year} {2009})}\BibitemShut
  {NoStop}%
\bibitem [{\citenamefont {Kotera}\ and\ \citenamefont {Olinto}(2011)}]{KO2011}%
  \BibitemOpen
  \bibfield  {author} {\bibinfo {author} {\bibfnamefont {K.}~\bibnamefont
  {Kotera}}\ and\ \bibinfo {author} {\bibfnamefont {A.~V.}\ \bibnamefont
  {Olinto}},\ }\bibfield  {title} {\bibinfo {title} {The astrophysics of
  ultrahigh-energy cosmic rays},\ }\href
  {https://doi.org/10.1146/annurev-astro-081710-102620} {\bibfield  {journal}
  {\bibinfo  {journal} {Annual Review of Astronomy and Astrophysics}\ }\textbf
  {\bibinfo {volume} {49}},\ \bibinfo {pages} {119} (\bibinfo {year}
  {2011})}\BibitemShut {NoStop}%
\bibitem [{\citenamefont {Alves~Batista}\ \emph {et~al.}(2019)\citenamefont
  {Alves~Batista}, \citenamefont {Biteau}, \citenamefont {Bustamante},
  \citenamefont {Dolag}, \citenamefont {Engel}, \citenamefont {Fang},
  \citenamefont {Kampert}, \citenamefont {Kostunin}, \citenamefont {Mostafa},
  \citenamefont {Murase},\ and\ \citenamefont {et~al.}}]{AB2019}%
  \BibitemOpen
  \bibfield  {author} {\bibinfo {author} {\bibfnamefont {R.}~\bibnamefont
  {Alves~Batista}}, \bibinfo {author} {\bibfnamefont {J.}~\bibnamefont
  {Biteau}}, \bibinfo {author} {\bibfnamefont {M.}~\bibnamefont {Bustamante}},
  \bibinfo {author} {\bibfnamefont {K.}~\bibnamefont {Dolag}}, \bibinfo
  {author} {\bibfnamefont {R.}~\bibnamefont {Engel}}, \bibinfo {author}
  {\bibfnamefont {K.}~\bibnamefont {Fang}}, \bibinfo {author} {\bibfnamefont
  {K.-H.}\ \bibnamefont {Kampert}}, \bibinfo {author} {\bibfnamefont
  {D.}~\bibnamefont {Kostunin}}, \bibinfo {author} {\bibfnamefont
  {M.}~\bibnamefont {Mostafa}}, \bibinfo {author} {\bibfnamefont
  {K.}~\bibnamefont {Murase}},\ and\ \bibinfo {author} {\bibnamefont
  {et~al.}},\ }\bibfield  {title} {\bibinfo {title} {Open questions in
  cosmic-ray research at ultrahigh energies},\ }\href
  {https://doi.org/10.3389/fspas.2019.00023} {\bibfield  {journal} {\bibinfo
  {journal} {Frontiers in Astronomy and Space Sciences}\ }\textbf {\bibinfo
  {volume} {6}},\ \bibinfo {pages} {23} (\bibinfo {year} {2019})}\BibitemShut
  {NoStop}%
\bibitem [{\citenamefont {Anchordoqui}(2019)}]{Luis2019}%
  \BibitemOpen
  \bibfield  {author} {\bibinfo {author} {\bibfnamefont {L.~A.}\ \bibnamefont
  {Anchordoqui}},\ }\bibfield  {title} {\bibinfo {title} {Ultra-high-energy
  cosmic rays},\ }\href {https://doi.org/10.1016/j.physrep.2019.01.002}
  {\bibfield  {journal} {\bibinfo  {journal} {Physics Reports}\ }\textbf
  {\bibinfo {volume} {801}},\ \bibinfo {pages} {1} (\bibinfo {year}
  {2019})}\BibitemShut {NoStop}%
\bibitem [{\citenamefont {Aab}\ \emph {et~al.}(2015)\citenamefont {Aab} \emph
  {et~al.}}]{PAO2015}%
  \BibitemOpen
  \bibfield  {author} {\bibinfo {author} {\bibfnamefont {A.}~\bibnamefont
  {Aab}} \emph {et~al.} (\bibinfo {collaboration} {Pierre Auger}),\ }\bibfield
  {title} {\bibinfo {title} {{The Pierre Auger Cosmic Ray Observatory}},\
  }\href {https://doi.org/10.1016/j.nima.2015.06.058} {\bibfield  {journal}
  {\bibinfo  {journal} {Nucl. Instrum. Meth. A}\ }\textbf {\bibinfo {volume}
  {798}},\ \bibinfo {pages} {172} (\bibinfo {year} {2015})},\ \Eprint
  {https://arxiv.org/abs/1502.01323} {arXiv:1502.01323 [astro-ph.IM]}
  \BibitemShut {NoStop}%
\bibitem [{\citenamefont {Abu-Zayyad}\ \emph {et~al.}(2013)\citenamefont
  {Abu-Zayyad}, \citenamefont {Aida}, \citenamefont {Allen}, \citenamefont
  {Anderson}, \citenamefont {Azuma}, \citenamefont {Barcikowski}, \citenamefont
  {Belz}, \citenamefont {Bergman}, \citenamefont {Blake}, \citenamefont
  {Cady},\ and\ \citenamefont {et~al.}}]{AZ2013}%
  \BibitemOpen
  \bibfield  {author} {\bibinfo {author} {\bibfnamefont {T.}~\bibnamefont
  {Abu-Zayyad}}, \bibinfo {author} {\bibfnamefont {R.}~\bibnamefont {Aida}},
  \bibinfo {author} {\bibfnamefont {M.}~\bibnamefont {Allen}}, \bibinfo
  {author} {\bibfnamefont {R.}~\bibnamefont {Anderson}}, \bibinfo {author}
  {\bibfnamefont {R.}~\bibnamefont {Azuma}}, \bibinfo {author} {\bibfnamefont
  {E.}~\bibnamefont {Barcikowski}}, \bibinfo {author} {\bibfnamefont {J.~W.}\
  \bibnamefont {Belz}}, \bibinfo {author} {\bibfnamefont {D.~R.}\ \bibnamefont
  {Bergman}}, \bibinfo {author} {\bibfnamefont {S.~A.}\ \bibnamefont {Blake}},
  \bibinfo {author} {\bibfnamefont {R.}~\bibnamefont {Cady}},\ and\ \bibinfo
  {author} {\bibnamefont {et~al.}},\ }\bibfield  {title} {\bibinfo {title} {The
  cosmic-ray energy spectrum observed with the surface detector of the
  telescope array experiment},\ }\href
  {https://doi.org/10.1088/2041-8205/768/1/l1} {\bibfield  {journal} {\bibinfo
  {journal} {The Astrophysical Journal}\ }\textbf {\bibinfo {volume} {768}},\
  \bibinfo {pages} {L1} (\bibinfo {year} {2013})}\BibitemShut {NoStop}%
\bibitem [{\citenamefont {Greisen}(1966)}]{Greisen:1966jv}%
  \BibitemOpen
  \bibfield  {author} {\bibinfo {author} {\bibfnamefont {K.}~\bibnamefont
  {Greisen}},\ }\bibfield  {title} {\bibinfo {title} {{End to the cosmic ray
  spectrum?}},\ }\href {https://doi.org/10.1103/PhysRevLett.16.748} {\bibfield
  {journal} {\bibinfo  {journal} {Phys. Rev. Lett.}\ }\textbf {\bibinfo
  {volume} {16}},\ \bibinfo {pages} {748} (\bibinfo {year} {1966})}\BibitemShut
  {NoStop}%
\bibitem [{\citenamefont {{Zatsepin}}\ and\ \citenamefont
  {{Kuz'min}}(1966)}]{1966JETPL...4...78Z}%
  \BibitemOpen
  \bibfield  {author} {\bibinfo {author} {\bibfnamefont {G.~T.}\ \bibnamefont
  {{Zatsepin}}}\ and\ \bibinfo {author} {\bibfnamefont {V.~A.}\ \bibnamefont
  {{Kuz'min}}},\ }\bibfield  {title} {\bibinfo {title} {{Upper Limit of the
  Spectrum of Cosmic Rays}},\ }\href@noop {} {\bibfield  {journal} {\bibinfo
  {journal} {Soviet Journal of Experimental and Theoretical Physics Letters}\
  }\textbf {\bibinfo {volume} {4}},\ \bibinfo {pages} {78} (\bibinfo {year}
  {1966})}\BibitemShut {NoStop}%
\bibitem [{\citenamefont {Abraham}\ \emph {et~al.}(2008)\citenamefont {Abraham}
  \emph {et~al.}}]{Abraham2008}%
  \BibitemOpen
  \bibfield  {author} {\bibinfo {author} {\bibfnamefont {J.}~\bibnamefont
  {Abraham}} \emph {et~al.} (\bibinfo {collaboration} {Pierre Auger}),\
  }\bibfield  {title} {\bibinfo {title} {{Observation of the suppression of the
  flux of cosmic rays above $4\times 10^{19}$eV}},\ }\href
  {https://doi.org/10.1103/PhysRevLett.101.061101} {\bibfield  {journal}
  {\bibinfo  {journal} {Phys. Rev. Lett.}\ }\textbf {\bibinfo {volume} {101}},\
  \bibinfo {pages} {061101} (\bibinfo {year} {2008})},\ \Eprint
  {https://arxiv.org/abs/0806.4302} {arXiv:0806.4302 [astro-ph]} \BibitemShut
  {NoStop}%
\bibitem [{\citenamefont {Abbasi}\ \emph {et~al.}(2008)\citenamefont {Abbasi},
  \citenamefont {Abu-Zayyad}, \citenamefont {Allen}, \citenamefont {Amman},
  \citenamefont {Archbold}, \citenamefont {Belov}, \citenamefont {Belz},
  \citenamefont {Ben~Zvi}, \citenamefont {Bergman}, \citenamefont {Blake},
  \citenamefont {Brusova}, \citenamefont {Burt}, \citenamefont {Cannon},
  \citenamefont {Cao}, \citenamefont {Connolly}, \citenamefont {Deng},
  \citenamefont {Fedorova}, \citenamefont {Finley}, \citenamefont {Gray},
  \citenamefont {Hanlon}, \citenamefont {Hoffman}, \citenamefont
  {Holzscheiter}, \citenamefont {Hughes}, \citenamefont {H\"untemeyer},
  \citenamefont {Jones}, \citenamefont {Jui}, \citenamefont {Kim},
  \citenamefont {Kirn}, \citenamefont {Loh}, \citenamefont {Maestas},
  \citenamefont {Manago}, \citenamefont {Marek}, \citenamefont {Martens},
  \citenamefont {Matthews}, \citenamefont {Matthews}, \citenamefont {Moore},
  \citenamefont {O'Neill}, \citenamefont {Painter}, \citenamefont {Perera},
  \citenamefont {Reil}, \citenamefont {Riehle}, \citenamefont {Roberts},
  \citenamefont {Rodriguez}, \citenamefont {Sasaki}, \citenamefont {Schnetzer},
  \citenamefont {Scott}, \citenamefont {Sinnis}, \citenamefont {Smith},
  \citenamefont {Sokolsky}, \citenamefont {Song}, \citenamefont {Springer},
  \citenamefont {Stokes}, \citenamefont {Thomas}, \citenamefont {Thomas},
  \citenamefont {Thomson}, \citenamefont {Tupa}, \citenamefont {Westerhoff},
  \citenamefont {Wiencke}, \citenamefont {Zhang},\ and\ \citenamefont
  {Zech}}]{Abbasi2008}%
  \BibitemOpen
  \bibfield  {author} {\bibinfo {author} {\bibfnamefont {R.~U.}\ \bibnamefont
  {Abbasi}}, \bibinfo {author} {\bibfnamefont {T.}~\bibnamefont {Abu-Zayyad}},
  \bibinfo {author} {\bibfnamefont {M.}~\bibnamefont {Allen}}, \bibinfo
  {author} {\bibfnamefont {J.~F.}\ \bibnamefont {Amman}}, \bibinfo {author}
  {\bibfnamefont {G.}~\bibnamefont {Archbold}}, \bibinfo {author}
  {\bibfnamefont {K.}~\bibnamefont {Belov}}, \bibinfo {author} {\bibfnamefont
  {J.~W.}\ \bibnamefont {Belz}}, \bibinfo {author} {\bibfnamefont {S.~Y.}\
  \bibnamefont {Ben~Zvi}}, \bibinfo {author} {\bibfnamefont {D.~R.}\
  \bibnamefont {Bergman}}, \bibinfo {author} {\bibfnamefont {S.~A.}\
  \bibnamefont {Blake}}, \bibinfo {author} {\bibfnamefont {O.~A.}\ \bibnamefont
  {Brusova}}, \bibinfo {author} {\bibfnamefont {G.~W.}\ \bibnamefont {Burt}},
  \bibinfo {author} {\bibfnamefont {C.}~\bibnamefont {Cannon}}, \bibinfo
  {author} {\bibfnamefont {Z.}~\bibnamefont {Cao}}, \bibinfo {author}
  {\bibfnamefont {B.~C.}\ \bibnamefont {Connolly}}, \bibinfo {author}
  {\bibfnamefont {W.}~\bibnamefont {Deng}}, \bibinfo {author} {\bibfnamefont
  {Y.}~\bibnamefont {Fedorova}}, \bibinfo {author} {\bibfnamefont {C.~B.}\
  \bibnamefont {Finley}}, \bibinfo {author} {\bibfnamefont {R.~C.}\
  \bibnamefont {Gray}}, \bibinfo {author} {\bibfnamefont {W.~F.}\ \bibnamefont
  {Hanlon}}, \bibinfo {author} {\bibfnamefont {C.~M.}\ \bibnamefont {Hoffman}},
  \bibinfo {author} {\bibfnamefont {M.~H.}\ \bibnamefont {Holzscheiter}},
  \bibinfo {author} {\bibfnamefont {G.}~\bibnamefont {Hughes}}, \bibinfo
  {author} {\bibfnamefont {P.}~\bibnamefont {H\"untemeyer}}, \bibinfo {author}
  {\bibfnamefont {B.~F.}\ \bibnamefont {Jones}}, \bibinfo {author}
  {\bibfnamefont {C.~C.~H.}\ \bibnamefont {Jui}}, \bibinfo {author}
  {\bibfnamefont {K.}~\bibnamefont {Kim}}, \bibinfo {author} {\bibfnamefont
  {M.~A.}\ \bibnamefont {Kirn}}, \bibinfo {author} {\bibfnamefont {E.~C.}\
  \bibnamefont {Loh}}, \bibinfo {author} {\bibfnamefont {M.~M.}\ \bibnamefont
  {Maestas}}, \bibinfo {author} {\bibfnamefont {N.}~\bibnamefont {Manago}},
  \bibinfo {author} {\bibfnamefont {L.~J.}\ \bibnamefont {Marek}}, \bibinfo
  {author} {\bibfnamefont {K.}~\bibnamefont {Martens}}, \bibinfo {author}
  {\bibfnamefont {J.~A.~J.}\ \bibnamefont {Matthews}}, \bibinfo {author}
  {\bibfnamefont {J.~N.}\ \bibnamefont {Matthews}}, \bibinfo {author}
  {\bibfnamefont {S.~A.}\ \bibnamefont {Moore}}, \bibinfo {author}
  {\bibfnamefont {A.}~\bibnamefont {O'Neill}}, \bibinfo {author} {\bibfnamefont
  {C.~A.}\ \bibnamefont {Painter}}, \bibinfo {author} {\bibfnamefont
  {L.}~\bibnamefont {Perera}}, \bibinfo {author} {\bibfnamefont
  {K.}~\bibnamefont {Reil}}, \bibinfo {author} {\bibfnamefont {R.}~\bibnamefont
  {Riehle}}, \bibinfo {author} {\bibfnamefont {M.}~\bibnamefont {Roberts}},
  \bibinfo {author} {\bibfnamefont {D.}~\bibnamefont {Rodriguez}}, \bibinfo
  {author} {\bibfnamefont {N.}~\bibnamefont {Sasaki}}, \bibinfo {author}
  {\bibfnamefont {S.~R.}\ \bibnamefont {Schnetzer}}, \bibinfo {author}
  {\bibfnamefont {L.~M.}\ \bibnamefont {Scott}}, \bibinfo {author}
  {\bibfnamefont {G.}~\bibnamefont {Sinnis}}, \bibinfo {author} {\bibfnamefont
  {J.~D.}\ \bibnamefont {Smith}}, \bibinfo {author} {\bibfnamefont
  {P.}~\bibnamefont {Sokolsky}}, \bibinfo {author} {\bibfnamefont
  {C.}~\bibnamefont {Song}}, \bibinfo {author} {\bibfnamefont {R.~W.}\
  \bibnamefont {Springer}}, \bibinfo {author} {\bibfnamefont {B.~T.}\
  \bibnamefont {Stokes}}, \bibinfo {author} {\bibfnamefont {S.~B.}\
  \bibnamefont {Thomas}}, \bibinfo {author} {\bibfnamefont {J.~R.}\
  \bibnamefont {Thomas}}, \bibinfo {author} {\bibfnamefont {G.~B.}\
  \bibnamefont {Thomson}}, \bibinfo {author} {\bibfnamefont {D.}~\bibnamefont
  {Tupa}}, \bibinfo {author} {\bibfnamefont {S.}~\bibnamefont {Westerhoff}},
  \bibinfo {author} {\bibfnamefont {L.~R.}\ \bibnamefont {Wiencke}}, \bibinfo
  {author} {\bibfnamefont {X.}~\bibnamefont {Zhang}},\ and\ \bibinfo {author}
  {\bibfnamefont {A.}~\bibnamefont {Zech}} (\bibinfo {collaboration} {High
  Resolution Fly's Eye Collaboration}),\ }\bibfield  {title} {\bibinfo {title}
  {First observation of the greisen-zatsepin-kuzmin suppression},\ }\href
  {https://doi.org/10.1103/PhysRevLett.100.101101} {\bibfield  {journal}
  {\bibinfo  {journal} {Phys. Rev. Lett.}\ }\textbf {\bibinfo {volume} {100}},\
  \bibinfo {pages} {101101} (\bibinfo {year} {2008})}\BibitemShut {NoStop}%
\bibitem [{\citenamefont {Halim}\ \emph {et~al.}(2023)\citenamefont {Halim}
  \emph {et~al.}}]{PierreAuger:2022atd}%
  \BibitemOpen
  \bibfield  {author} {\bibinfo {author} {\bibfnamefont {A.~A.}\ \bibnamefont
  {Halim}} \emph {et~al.} (\bibinfo {collaboration} {Pierre Auger}),\
  }\bibfield  {title} {\bibinfo {title} {{Constraining the sources of
  ultra-high-energy cosmic rays across and above the ankle with the spectrum
  and composition data measured at the Pierre Auger Observatory}},\ }\href
  {https://doi.org/10.1088/1475-7516/2023/05/024} {\bibfield  {journal}
  {\bibinfo  {journal} {JCAP}\ }\textbf {\bibinfo {volume} {05}},\ \bibinfo
  {pages} {024}},\ \Eprint {https://arxiv.org/abs/2211.02857} {arXiv:2211.02857
  [astro-ph.HE]} \BibitemShut {NoStop}%
\bibitem [{\citenamefont {Anchordoqui}\ \emph {et~al.}(2000)\citenamefont
  {Anchordoqui}, \citenamefont {Dova}, \citenamefont {McCauley}, \citenamefont
  {Reucroft},\ and\ \citenamefont {Swain}}]{Anchordoqui:1999aj}%
  \BibitemOpen
  \bibfield  {author} {\bibinfo {author} {\bibfnamefont {L.~A.}\ \bibnamefont
  {Anchordoqui}}, \bibinfo {author} {\bibfnamefont {M.~T.}\ \bibnamefont
  {Dova}}, \bibinfo {author} {\bibfnamefont {T.~P.}\ \bibnamefont {McCauley}},
  \bibinfo {author} {\bibfnamefont {S.}~\bibnamefont {Reucroft}},\ and\
  \bibinfo {author} {\bibfnamefont {J.~D.}\ \bibnamefont {Swain}},\ }\bibfield
  {title} {\bibinfo {title} {{Possible explanation for the tail of the cosmic
  ray spectrum}},\ }\href {https://doi.org/10.1016/S0370-2693(00)00534-7}
  {\bibfield  {journal} {\bibinfo  {journal} {Phys. Lett. B}\ }\textbf
  {\bibinfo {volume} {482}},\ \bibinfo {pages} {343} (\bibinfo {year}
  {2000})},\ \Eprint {https://arxiv.org/abs/astro-ph/9912081}
  {arXiv:astro-ph/9912081} \BibitemShut {NoStop}%
\bibitem [{\citenamefont {Zhang}\ \emph {et~al.}(2026)\citenamefont {Zhang},
  \citenamefont {Murase}, \citenamefont {Ekanger}, \citenamefont
  {Bhattacharya},\ and\ \citenamefont {Horiuchi}}]{Zhang:2024sjp}%
  \BibitemOpen
  \bibfield  {author} {\bibinfo {author} {\bibfnamefont {B.~T.}\ \bibnamefont
  {Zhang}}, \bibinfo {author} {\bibfnamefont {K.}~\bibnamefont {Murase}},
  \bibinfo {author} {\bibfnamefont {N.}~\bibnamefont {Ekanger}}, \bibinfo
  {author} {\bibfnamefont {M.}~\bibnamefont {Bhattacharya}},\ and\ \bibinfo
  {author} {\bibfnamefont {S.}~\bibnamefont {Horiuchi}},\ }\bibfield  {title}
  {\bibinfo {title} {{Ultraheavy Ultrahigh-Energy Cosmic Rays}},\ }\href
  {https://doi.org/10.1103/221m-gvs3} {\bibfield  {journal} {\bibinfo
  {journal} {Phys. Rev. Lett.}\ }\textbf {\bibinfo {volume} {136}},\ \bibinfo
  {pages} {181002} (\bibinfo {year} {2026})},\ \Eprint
  {https://arxiv.org/abs/2405.17409} {arXiv:2405.17409 [astro-ph.HE]}
  \BibitemShut {NoStop}%
\bibitem [{\citenamefont {{Jiang}}\ \emph {et~al.}(2021)\citenamefont
  {{Jiang}}, \citenamefont {{Zhang}},\ and\ \citenamefont
  {{Murase}}}]{Jiang2021}%
  \BibitemOpen
  \bibfield  {author} {\bibinfo {author} {\bibfnamefont {Y.}~\bibnamefont
  {{Jiang}}}, \bibinfo {author} {\bibfnamefont {B.~T.}\ \bibnamefont
  {{Zhang}}},\ and\ \bibinfo {author} {\bibfnamefont {K.}~\bibnamefont
  {{Murase}}},\ }\bibfield  {title} {\bibinfo {title} {Energetics of
  ultrahigh-energy cosmic-ray nuclei},\ }\href
  {https://doi.org/10.1103/PhysRevD.104.043017} {\bibfield  {journal} {\bibinfo
   {journal} {Phys. Rev. D}\ }\textbf {\bibinfo {volume} {104}},\ \bibinfo
  {pages} {043017} (\bibinfo {year} {2021})}\BibitemShut {NoStop}%
\bibitem [{\citenamefont {Kuznetsov}\ and\ \citenamefont
  {Tinyakov}(2021)}]{KT2021}%
  \BibitemOpen
  \bibfield  {author} {\bibinfo {author} {\bibfnamefont {M.}~\bibnamefont
  {Kuznetsov}}\ and\ \bibinfo {author} {\bibfnamefont {P.}~\bibnamefont
  {Tinyakov}},\ }\bibfield  {title} {\bibinfo {title} {Uhecr mass composition
  at highest energies from anisotropy of their arrival directions},\ }\href
  {https://doi.org/10.1088/1475-7516/2021/04/065} {\bibfield  {journal}
  {\bibinfo  {journal} {Journal of Cosmology and Astroparticle Physics}\
  }\textbf {\bibinfo {volume} {2021}}\bibinfo  {number} { (04)},\ \bibinfo
  {pages} {065}}\BibitemShut {NoStop}%
\bibitem [{\citenamefont {{Norman}}\ \emph {et~al.}(1995)\citenamefont
  {{Norman}}, \citenamefont {{Melrose}},\ and\ \citenamefont
  {{Achterberg}}}]{Norman1995}%
  \BibitemOpen
\bibfield  {number} {  }\bibfield  {author} {\bibinfo {author} {\bibfnamefont
  {C.~A.}\ \bibnamefont {{Norman}}}, \bibinfo {author} {\bibfnamefont {D.~B.}\
  \bibnamefont {{Melrose}}},\ and\ \bibinfo {author} {\bibfnamefont
  {A.}~\bibnamefont {{Achterberg}}},\ }\bibfield  {title} {\bibinfo {title}
  {The origin of cosmic rays above 10 18.5 ev},\ }\href
  {https://doi.org/10.1086/176465} {\bibfield  {journal} {\bibinfo  {journal}
  {The Astrophysical Journal}\ }\textbf {\bibinfo {volume} {454}},\ \bibinfo
  {pages} {60} (\bibinfo {year} {1995})}\BibitemShut {NoStop}%
\bibitem [{\citenamefont {{Dermer}}\ \emph {et~al.}(2009)\citenamefont
  {{Dermer}}, \citenamefont {{Razzaque}}, \citenamefont {{Finke}},\ and\
  \citenamefont {{Atoyan}}}]{Dermer2009}%
  \BibitemOpen
  \bibfield  {author} {\bibinfo {author} {\bibfnamefont {C.~D.}\ \bibnamefont
  {{Dermer}}}, \bibinfo {author} {\bibfnamefont {S.}~\bibnamefont
  {{Razzaque}}}, \bibinfo {author} {\bibfnamefont {J.~D.}\ \bibnamefont
  {{Finke}}},\ and\ \bibinfo {author} {\bibfnamefont {A.}~\bibnamefont
  {{Atoyan}}},\ }\bibfield  {title} {\bibinfo {title} {Ultra-high-energy cosmic
  rays from black hole jets of radio galaxies},\ }\href
  {https://doi.org/10.1088/1367-2630/11/6/065016} {\bibfield  {journal}
  {\bibinfo  {journal} {New J. Phys.}\ }\textbf {\bibinfo {volume} {11}},\
  \bibinfo {pages} {065016} (\bibinfo {year} {2009})}\BibitemShut {NoStop}%
\bibitem [{\citenamefont {Pe'er}\ \emph {et~al.}(2009)\citenamefont {Pe'er},
  \citenamefont {Murase},\ and\ \citenamefont {Meszaros}}]{Peer:2009vnw}%
  \BibitemOpen
  \bibfield  {author} {\bibinfo {author} {\bibfnamefont {A.}~\bibnamefont
  {Pe'er}}, \bibinfo {author} {\bibfnamefont {K.}~\bibnamefont {Murase}},\ and\
  \bibinfo {author} {\bibfnamefont {P.}~\bibnamefont {Meszaros}},\ }\bibfield
  {title} {\bibinfo {title} {{Radio Quiet AGNs as Possible Sources of
  Ultra-high Energy Cosmic Rays}},\ }\href
  {https://doi.org/10.1103/PhysRevD.80.123018} {\bibfield  {journal} {\bibinfo
  {journal} {Phys. Rev. D}\ }\textbf {\bibinfo {volume} {80}},\ \bibinfo
  {pages} {123018} (\bibinfo {year} {2009})},\ \Eprint
  {https://arxiv.org/abs/0911.1776} {arXiv:0911.1776 [astro-ph.HE]}
  \BibitemShut {NoStop}%
\bibitem [{\citenamefont {{Takami}}\ and\ \citenamefont
  {{Horiuchi}}(2011)}]{TH2011}%
  \BibitemOpen
  \bibfield  {author} {\bibinfo {author} {\bibfnamefont {H.}~\bibnamefont
  {{Takami}}}\ and\ \bibinfo {author} {\bibfnamefont {S.}~\bibnamefont
  {{Horiuchi}}},\ }\bibfield  {title} {\bibinfo {title} {The production of
  ultra high energy cosmic rays during the early epochs of radio-loud agn},\
  }\href {https://doi.org/10.1016/j.astropartphys.2011.01.014} {\bibfield
  {journal} {\bibinfo  {journal} {Astroparticle Physics}\ }\textbf {\bibinfo
  {volume} {34}},\ \bibinfo {pages} {749} (\bibinfo {year} {2011})}\BibitemShut
  {NoStop}%
\bibitem [{\citenamefont {Murase}\ \emph {et~al.}(2012)\citenamefont {Murase},
  \citenamefont {Dermer}, \citenamefont {Takami},\ and\ \citenamefont
  {Migliori}}]{Murase:2011cy}%
  \BibitemOpen
  \bibfield  {author} {\bibinfo {author} {\bibfnamefont {K.}~\bibnamefont
  {Murase}}, \bibinfo {author} {\bibfnamefont {C.~D.}\ \bibnamefont {Dermer}},
  \bibinfo {author} {\bibfnamefont {H.}~\bibnamefont {Takami}},\ and\ \bibinfo
  {author} {\bibfnamefont {G.}~\bibnamefont {Migliori}},\ }\bibfield  {title}
  {\bibinfo {title} {{Blazars as Ultra-High-Energy Cosmic-Ray Sources:
  Implications for TeV Gamma-Ray Observations}},\ }\href
  {https://doi.org/10.1088/0004-637X/749/1/63} {\bibfield  {journal} {\bibinfo
  {journal} {Astrophys. J.}\ }\textbf {\bibinfo {volume} {749}},\ \bibinfo
  {pages} {63} (\bibinfo {year} {2012})},\ \Eprint
  {https://arxiv.org/abs/1107.5576} {arXiv:1107.5576 [astro-ph.HE]}
  \BibitemShut {NoStop}%
\bibitem [{\citenamefont {{Waxman}}(1995)}]{Waxman1995}%
  \BibitemOpen
  \bibfield  {author} {\bibinfo {author} {\bibfnamefont {E.}~\bibnamefont
  {{Waxman}}},\ }\bibfield  {title} {\bibinfo {title} {Cosmological gamma-ray
  bursts and the highest energy cosmic rays},\ }\href
  {https://doi.org/10.1103/PhysRevLett.75.386} {\bibfield  {journal} {\bibinfo
  {journal} {Phys. Rev. Lett.}\ }\textbf {\bibinfo {volume} {75}},\ \bibinfo
  {pages} {386} (\bibinfo {year} {1995})}\BibitemShut {NoStop}%
\bibitem [{\citenamefont {{Milgrom}}\ and\ \citenamefont
  {{Usov}}(1995)}]{MU1995}%
  \BibitemOpen
  \bibfield  {author} {\bibinfo {author} {\bibfnamefont {M.}~\bibnamefont
  {{Milgrom}}}\ and\ \bibinfo {author} {\bibfnamefont {V.}~\bibnamefont
  {{Usov}}},\ }\bibfield  {title} {\bibinfo {title} {Ascertaining the core
  collapse supernova mechanism: The state of the art and the road ahead},\
  }\href {https://doi.org/10.1086/309633} {\bibfield  {journal} {\bibinfo
  {journal} {ApJL}\ }\textbf {\bibinfo {volume} {449}},\ \bibinfo {pages} {L37}
  (\bibinfo {year} {1995})}\BibitemShut {NoStop}%
\bibitem [{\citenamefont {Murase}\ \emph {et~al.}(2006)\citenamefont {Murase},
  \citenamefont {Ioka}, \citenamefont {Nagataki},\ and\ \citenamefont
  {Nakamura}}]{Murase2006}%
  \BibitemOpen
  \bibfield  {author} {\bibinfo {author} {\bibfnamefont {K.}~\bibnamefont
  {Murase}}, \bibinfo {author} {\bibfnamefont {K.}~\bibnamefont {Ioka}},
  \bibinfo {author} {\bibfnamefont {S.}~\bibnamefont {Nagataki}},\ and\
  \bibinfo {author} {\bibfnamefont {T.}~\bibnamefont {Nakamura}},\ }\bibfield
  {title} {\bibinfo {title} {{High Energy Neutrinos and Cosmic-Rays from
  Low-Luminosity Gamma-Ray Bursts?}},\ }\href {https://doi.org/10.1086/509323}
  {\bibfield  {journal} {\bibinfo  {journal} {Astrophys. J. Lett.}\ }\textbf
  {\bibinfo {volume} {651}},\ \bibinfo {pages} {L5} (\bibinfo {year} {2006})},\
  \Eprint {https://arxiv.org/abs/astro-ph/0607104} {arXiv:astro-ph/0607104}
  \BibitemShut {NoStop}%
\bibitem [{\citenamefont {Murase}\ \emph {et~al.}(2008)\citenamefont {Murase},
  \citenamefont {Ioka}, \citenamefont {Nagataki},\ and\ \citenamefont
  {Nakamura}}]{Murase:2008mr}%
  \BibitemOpen
  \bibfield  {author} {\bibinfo {author} {\bibfnamefont {K.}~\bibnamefont
  {Murase}}, \bibinfo {author} {\bibfnamefont {K.}~\bibnamefont {Ioka}},
  \bibinfo {author} {\bibfnamefont {S.}~\bibnamefont {Nagataki}},\ and\
  \bibinfo {author} {\bibfnamefont {T.}~\bibnamefont {Nakamura}},\ }\bibfield
  {title} {\bibinfo {title} {{High-energy cosmic-ray nuclei from high- and
  low-luminosity gamma-ray bursts and implications for multi-messenger
  astronomy}},\ }\href {https://doi.org/10.1103/PhysRevD.78.023005} {\bibfield
  {journal} {\bibinfo  {journal} {Phys. Rev. D}\ }\textbf {\bibinfo {volume}
  {78}},\ \bibinfo {pages} {023005} (\bibinfo {year} {2008})},\ \Eprint
  {https://arxiv.org/abs/0801.2861} {arXiv:0801.2861 [astro-ph]} \BibitemShut
  {NoStop}%
\bibitem [{\citenamefont {Liu}\ and\ \citenamefont {Wang}(2012)}]{Liu:2012zzd}%
  \BibitemOpen
  \bibfield  {author} {\bibinfo {author} {\bibfnamefont {R.-Y.}\ \bibnamefont
  {Liu}}\ and\ \bibinfo {author} {\bibfnamefont {X.-Y.}\ \bibnamefont {Wang}},\
  }\bibfield  {title} {\bibinfo {title} {{Energy spectrum and chemical
  composition of ultrahigh energy cosmic rays from semi-relativistic
  hypernovae}},\ }\href {https://doi.org/10.1088/0004-637X/746/1/40} {\bibfield
   {journal} {\bibinfo  {journal} {Astrophys. J.}\ }\textbf {\bibinfo {volume}
  {746}},\ \bibinfo {pages} {40} (\bibinfo {year} {2012})},\ \Eprint
  {https://arxiv.org/abs/1111.6256} {arXiv:1111.6256 [astro-ph.HE]}
  \BibitemShut {NoStop}%
\bibitem [{\citenamefont {Arons}(2003)}]{Arons2003}%
  \BibitemOpen
  \bibfield  {author} {\bibinfo {author} {\bibfnamefont {J.}~\bibnamefont
  {Arons}},\ }\bibfield  {title} {\bibinfo {title} {Magnetars in the
  metagalaxy: An origin for ultra–high‐energy cosmic rays in the nearby
  universe},\ }\href {https://doi.org/10.1086/374776} {\bibfield  {journal}
  {\bibinfo  {journal} {The Astrophysical Journal}\ }\textbf {\bibinfo {volume}
  {589}},\ \bibinfo {pages} {871} (\bibinfo {year} {2003})}\BibitemShut
  {NoStop}%
\bibitem [{\citenamefont {{Murase}}\ \emph {et~al.}(2009)\citenamefont
  {{Murase}}, \citenamefont {{Meszaros}},\ and\ \citenamefont
  {{Zhang}}}]{KM2009}%
  \BibitemOpen
  \bibfield  {author} {\bibinfo {author} {\bibfnamefont {K.}~\bibnamefont
  {{Murase}}}, \bibinfo {author} {\bibfnamefont {P.}~\bibnamefont
  {{Meszaros}}},\ and\ \bibinfo {author} {\bibfnamefont {B.}~\bibnamefont
  {{Zhang}}},\ }\bibfield  {title} {\bibinfo {title} {Probing the birth of fast
  rotating magnetars through high-energy neutrinos},\ }\href
  {https://doi.org/10.1103/PhysRevD.79.103001} {\bibfield  {journal} {\bibinfo
  {journal} {Phys. Rev. D.}\ }\textbf {\bibinfo {volume} {79}},\ \bibinfo
  {pages} {103001} (\bibinfo {year} {2009})}\BibitemShut {NoStop}%
\bibitem [{\citenamefont {Fang}\ \emph {et~al.}(2012)\citenamefont {Fang},
  \citenamefont {Kotera},\ and\ \citenamefont {Olinto}}]{Fang:2012rx}%
  \BibitemOpen
  \bibfield  {author} {\bibinfo {author} {\bibfnamefont {K.}~\bibnamefont
  {Fang}}, \bibinfo {author} {\bibfnamefont {K.}~\bibnamefont {Kotera}},\ and\
  \bibinfo {author} {\bibfnamefont {A.~V.}\ \bibnamefont {Olinto}},\ }\bibfield
   {title} {\bibinfo {title} {{Newly-born pulsars as sources of ultrahigh
  energy cosmic rays}},\ }\href {https://doi.org/10.1088/0004-637X/750/2/118}
  {\bibfield  {journal} {\bibinfo  {journal} {Astrophys. J.}\ }\textbf
  {\bibinfo {volume} {750}},\ \bibinfo {pages} {118} (\bibinfo {year}
  {2012})},\ \Eprint {https://arxiv.org/abs/1201.5197} {arXiv:1201.5197
  [astro-ph.HE]} \BibitemShut {NoStop}%
\bibitem [{\citenamefont {Fang}\ \emph {et~al.}(2014)\citenamefont {Fang},
  \citenamefont {Kotera}, \citenamefont {Murase},\ and\ \citenamefont
  {Olinto}}]{Fang:2013vla}%
  \BibitemOpen
  \bibfield  {author} {\bibinfo {author} {\bibfnamefont {K.}~\bibnamefont
  {Fang}}, \bibinfo {author} {\bibfnamefont {K.}~\bibnamefont {Kotera}},
  \bibinfo {author} {\bibfnamefont {K.}~\bibnamefont {Murase}},\ and\ \bibinfo
  {author} {\bibfnamefont {A.~V.}\ \bibnamefont {Olinto}},\ }\bibfield  {title}
  {\bibinfo {title} {{Testing the Newborn Pulsar Origin of Ultrahigh Energy
  Cosmic Rays with EeV Neutrinos}},\ }\href
  {https://doi.org/10.1103/PhysRevD.90.103005} {\bibfield  {journal} {\bibinfo
  {journal} {Phys. Rev. D}\ }\textbf {\bibinfo {volume} {90}},\ \bibinfo
  {pages} {103005} (\bibinfo {year} {2014})},\ \bibinfo {note} {[Erratum:
  Phys.Rev.D 92, 129901 (2015)]},\ \Eprint {https://arxiv.org/abs/1311.2044}
  {arXiv:1311.2044 [astro-ph.HE]} \BibitemShut {NoStop}%
\bibitem [{\citenamefont {Takami}\ \emph {et~al.}(2014)\citenamefont {Takami},
  \citenamefont {Kyutoku},\ and\ \citenamefont {Ioka}}]{Takami:2013rza}%
  \BibitemOpen
  \bibfield  {author} {\bibinfo {author} {\bibfnamefont {H.}~\bibnamefont
  {Takami}}, \bibinfo {author} {\bibfnamefont {K.}~\bibnamefont {Kyutoku}},\
  and\ \bibinfo {author} {\bibfnamefont {K.}~\bibnamefont {Ioka}},\ }\bibfield
  {title} {\bibinfo {title} {{High-Energy Radiation from Remnants of Neutron
  Star Binary Mergers}},\ }\href {https://doi.org/10.1103/PhysRevD.89.063006}
  {\bibfield  {journal} {\bibinfo  {journal} {Phys. Rev. D}\ }\textbf {\bibinfo
  {volume} {89}},\ \bibinfo {pages} {063006} (\bibinfo {year} {2014})},\
  \Eprint {https://arxiv.org/abs/1307.6805} {arXiv:1307.6805 [astro-ph.HE]}
  \BibitemShut {NoStop}%
\bibitem [{\citenamefont {Rodrigues}\ \emph {et~al.}(2019)\citenamefont
  {Rodrigues}, \citenamefont {Biehl}, \citenamefont {Boncioli},\ and\
  \citenamefont {Taylor}}]{Rodrigues:2018bjg}%
  \BibitemOpen
  \bibfield  {author} {\bibinfo {author} {\bibfnamefont {X.}~\bibnamefont
  {Rodrigues}}, \bibinfo {author} {\bibfnamefont {D.}~\bibnamefont {Biehl}},
  \bibinfo {author} {\bibfnamefont {D.}~\bibnamefont {Boncioli}},\ and\
  \bibinfo {author} {\bibfnamefont {A.~M.}\ \bibnamefont {Taylor}},\ }\bibfield
   {title} {\bibinfo {title} {{Binary neutron star merger remnants as sources
  of cosmic rays below the {\textquotedblleft}Ankle{\textquotedblright}}},\
  }\href {https://doi.org/10.1016/j.astropartphys.2018.10.007} {\bibfield
  {journal} {\bibinfo  {journal} {Astropart. Phys.}\ }\textbf {\bibinfo
  {volume} {106}},\ \bibinfo {pages} {10} (\bibinfo {year} {2019})},\ \Eprint
  {https://arxiv.org/abs/1806.01624} {arXiv:1806.01624 [astro-ph.HE]}
  \BibitemShut {NoStop}%
\bibitem [{\citenamefont {Murase}\ and\ \citenamefont
  {Fukugita}(2019)}]{Murase:2018utn}%
  \BibitemOpen
  \bibfield  {author} {\bibinfo {author} {\bibfnamefont {K.}~\bibnamefont
  {Murase}}\ and\ \bibinfo {author} {\bibfnamefont {M.}~\bibnamefont
  {Fukugita}},\ }\bibfield  {title} {\bibinfo {title} {{Energetics of
  High-Energy Cosmic Radiations}},\ }\href
  {https://doi.org/10.1103/PhysRevD.99.063012} {\bibfield  {journal} {\bibinfo
  {journal} {Phys. Rev. D}\ }\textbf {\bibinfo {volume} {99}},\ \bibinfo
  {pages} {063012} (\bibinfo {year} {2019})},\ \Eprint
  {https://arxiv.org/abs/1806.04194} {arXiv:1806.04194 [astro-ph.HE]}
  \BibitemShut {NoStop}%
\bibitem [{\citenamefont {Farrar}(2025)}]{Farrar:2024zsm}%
  \BibitemOpen
  \bibfield  {author} {\bibinfo {author} {\bibfnamefont {G.~R.}\ \bibnamefont
  {Farrar}},\ }\bibfield  {title} {\bibinfo {title} {{Binary Neutron Star
  Mergers as the Source of the Highest Energy Cosmic Rays}},\ }\href
  {https://doi.org/10.1103/PhysRevLett.134.081003} {\bibfield  {journal}
  {\bibinfo  {journal} {Phys. Rev. Lett.}\ }\textbf {\bibinfo {volume} {134}},\
  \bibinfo {pages} {081003} (\bibinfo {year} {2025})},\ \Eprint
  {https://arxiv.org/abs/2405.12004} {arXiv:2405.12004 [astro-ph.HE]}
  \BibitemShut {NoStop}%
\bibitem [{\citenamefont {Caprioli}(2015)}]{Caprioli:2015zka}%
  \BibitemOpen
  \bibfield  {author} {\bibinfo {author} {\bibfnamefont {D.}~\bibnamefont
  {Caprioli}},\ }\bibfield  {title} {\bibinfo {title} {{''Espresso''
  Acceleration of Ultra-high-energy Cosmic Rays}},\ }\href
  {https://doi.org/10.1088/2041-8205/811/2/L38} {\bibfield  {journal} {\bibinfo
   {journal} {Astrophys. J. Lett.}\ }\textbf {\bibinfo {volume} {811}},\
  \bibinfo {pages} {L38} (\bibinfo {year} {2015})},\ \Eprint
  {https://arxiv.org/abs/1505.06739} {arXiv:1505.06739 [astro-ph.HE]}
  \BibitemShut {NoStop}%
\bibitem [{\citenamefont {Kimura}\ \emph {et~al.}(2018)\citenamefont {Kimura},
  \citenamefont {Murase},\ and\ \citenamefont {Zhang}}]{Kimura:2017ubz}%
  \BibitemOpen
  \bibfield  {author} {\bibinfo {author} {\bibfnamefont {S.~S.}\ \bibnamefont
  {Kimura}}, \bibinfo {author} {\bibfnamefont {K.}~\bibnamefont {Murase}},\
  and\ \bibinfo {author} {\bibfnamefont {B.~T.}\ \bibnamefont {Zhang}},\
  }\bibfield  {title} {\bibinfo {title} {{Ultrahigh-energy Cosmic-ray Nuclei
  from Black Hole Jets: Recycling Galactic Cosmic Rays through Shear
  Acceleration}},\ }\href {https://doi.org/10.1103/PhysRevD.97.023026}
  {\bibfield  {journal} {\bibinfo  {journal} {Phys. Rev. D}\ }\textbf {\bibinfo
  {volume} {97}},\ \bibinfo {pages} {023026} (\bibinfo {year} {2018})},\
  \Eprint {https://arxiv.org/abs/1705.05027} {arXiv:1705.05027 [astro-ph.HE]}
  \BibitemShut {NoStop}%
\bibitem [{\citenamefont {Mbarek}\ and\ \citenamefont
  {Caprioli}(2019)}]{Mbarek:2019glq}%
  \BibitemOpen
  \bibfield  {author} {\bibinfo {author} {\bibfnamefont {R.}~\bibnamefont
  {Mbarek}}\ and\ \bibinfo {author} {\bibfnamefont {D.}~\bibnamefont
  {Caprioli}},\ }\bibfield  {title} {\bibinfo {title} {{Bottom-up Acceleration
  of Ultra-High-Energy Cosmic Rays in the Jets of Active Galactic Nuclei}},\
  }\href {https://doi.org/10.3847/1538-4357/ab4a08} {\bibfield  {journal}
  {\bibinfo  {journal} {Astrophys. J.}\ }\textbf {\bibinfo {volume} {886}},\
  \bibinfo {pages} {8} (\bibinfo {year} {2019})},\ \Eprint
  {https://arxiv.org/abs/1904.02720} {arXiv:1904.02720 [astro-ph.HE]}
  \BibitemShut {NoStop}%
\bibitem [{\citenamefont {Mbarek}\ \emph {et~al.}(2023)\citenamefont {Mbarek},
  \citenamefont {Caprioli},\ and\ \citenamefont {Murase}}]{Mbarek:2022nat}%
  \BibitemOpen
  \bibfield  {author} {\bibinfo {author} {\bibfnamefont {R.}~\bibnamefont
  {Mbarek}}, \bibinfo {author} {\bibfnamefont {D.}~\bibnamefont {Caprioli}},\
  and\ \bibinfo {author} {\bibfnamefont {K.}~\bibnamefont {Murase}},\
  }\bibfield  {title} {\bibinfo {title} {{High-energy Neutrino Emission from
  Espresso-reaccelerated Ions in Jets of Active Galactic Nuclei}},\ }\href
  {https://doi.org/10.3847/1538-4357/aca481} {\bibfield  {journal} {\bibinfo
  {journal} {Astrophys. J.}\ }\textbf {\bibinfo {volume} {942}},\ \bibinfo
  {pages} {37} (\bibinfo {year} {2023})},\ \Eprint
  {https://arxiv.org/abs/2207.07130} {arXiv:2207.07130 [astro-ph.HE]}
  \BibitemShut {NoStop}%
\bibitem [{\citenamefont {Zhang}\ \emph {et~al.}(2018)\citenamefont {Zhang},
  \citenamefont {Murase}, \citenamefont {Kimura}, \citenamefont {Horiuchi},\
  and\ \citenamefont {M\'esz\'aros}}]{Zhang:2017moz}%
  \BibitemOpen
  \bibfield  {author} {\bibinfo {author} {\bibfnamefont {B.~T.}\ \bibnamefont
  {Zhang}}, \bibinfo {author} {\bibfnamefont {K.}~\bibnamefont {Murase}},
  \bibinfo {author} {\bibfnamefont {S.~S.}\ \bibnamefont {Kimura}}, \bibinfo
  {author} {\bibfnamefont {S.}~\bibnamefont {Horiuchi}},\ and\ \bibinfo
  {author} {\bibfnamefont {P.}~\bibnamefont {M\'esz\'aros}},\ }\bibfield
  {title} {\bibinfo {title} {{Low-luminosity gamma-ray bursts as the sources of
  ultrahigh-energy cosmic ray nuclei}},\ }\href
  {https://doi.org/10.1103/PhysRevD.97.083010} {\bibfield  {journal} {\bibinfo
  {journal} {Phys. Rev. D}\ }\textbf {\bibinfo {volume} {97}},\ \bibinfo
  {pages} {083010} (\bibinfo {year} {2018})},\ \Eprint
  {https://arxiv.org/abs/1712.09984} {arXiv:1712.09984 [astro-ph.HE]}
  \BibitemShut {NoStop}%
\bibitem [{\citenamefont {Boncioli}\ \emph {et~al.}(2019)\citenamefont
  {Boncioli}, \citenamefont {Biehl},\ and\ \citenamefont
  {Winter}}]{Boncioli:2018lrv}%
  \BibitemOpen
  \bibfield  {author} {\bibinfo {author} {\bibfnamefont {D.}~\bibnamefont
  {Boncioli}}, \bibinfo {author} {\bibfnamefont {D.}~\bibnamefont {Biehl}},\
  and\ \bibinfo {author} {\bibfnamefont {W.}~\bibnamefont {Winter}},\
  }\bibfield  {title} {\bibinfo {title} {{On the common origin of cosmic rays
  across the ankle and diffuse neutrinos at the highest energies from
  low-luminosity Gamma-Ray Bursts}},\ }\href
  {https://doi.org/10.3847/1538-4357/aafda7} {\bibfield  {journal} {\bibinfo
  {journal} {Astrophys. J.}\ }\textbf {\bibinfo {volume} {872}},\ \bibinfo
  {pages} {110} (\bibinfo {year} {2019})},\ \Eprint
  {https://arxiv.org/abs/1808.07481} {arXiv:1808.07481 [astro-ph.HE]}
  \BibitemShut {NoStop}%
\bibitem [{\citenamefont {Zhang}\ and\ \citenamefont
  {Murase}(2019)}]{Zhang:2018agl}%
  \BibitemOpen
  \bibfield  {author} {\bibinfo {author} {\bibfnamefont {B.~T.}\ \bibnamefont
  {Zhang}}\ and\ \bibinfo {author} {\bibfnamefont {K.}~\bibnamefont {Murase}},\
  }\bibfield  {title} {\bibinfo {title} {{Ultrahigh-energy cosmic-ray nuclei
  and neutrinos from engine-driven supernovae}},\ }\href
  {https://doi.org/10.1103/PhysRevD.100.103004} {\bibfield  {journal} {\bibinfo
   {journal} {Phys. Rev. D}\ }\textbf {\bibinfo {volume} {100}},\ \bibinfo
  {pages} {103004} (\bibinfo {year} {2019})},\ \Eprint
  {https://arxiv.org/abs/1812.10289} {arXiv:1812.10289 [astro-ph.HE]}
  \BibitemShut {NoStop}%
\bibitem [{\citenamefont {Zhang}\ \emph {et~al.}(2017)\citenamefont {Zhang},
  \citenamefont {Murase}, \citenamefont {Oikonomou},\ and\ \citenamefont
  {Li}}]{Zhang:2017hom}%
  \BibitemOpen
  \bibfield  {author} {\bibinfo {author} {\bibfnamefont {B.~T.}\ \bibnamefont
  {Zhang}}, \bibinfo {author} {\bibfnamefont {K.}~\bibnamefont {Murase}},
  \bibinfo {author} {\bibfnamefont {F.}~\bibnamefont {Oikonomou}},\ and\
  \bibinfo {author} {\bibfnamefont {Z.}~\bibnamefont {Li}},\ }\bibfield
  {title} {\bibinfo {title} {{High-energy cosmic ray nuclei from tidal
  disruption events: Origin, survival, and implications}},\ }\href
  {https://doi.org/10.1103/PhysRevD.96.063007} {\bibfield  {journal} {\bibinfo
  {journal} {Phys. Rev. D}\ }\textbf {\bibinfo {volume} {96}},\ \bibinfo
  {pages} {063007} (\bibinfo {year} {2017})},\ \bibinfo {note} {[Addendum:
  Phys.Rev.D 96, 069902 (2017)]},\ \Eprint {https://arxiv.org/abs/1706.00391}
  {arXiv:1706.00391 [astro-ph.HE]} \BibitemShut {NoStop}%
\bibitem [{\citenamefont {Metzger}\ \emph
  {et~al.}(2011{\natexlab{a}})\citenamefont {Metzger}, \citenamefont
  {Giannios},\ and\ \citenamefont {Horiuchi}}]{Metzger2011b}%
  \BibitemOpen
  \bibfield  {author} {\bibinfo {author} {\bibfnamefont {B.~D.}\ \bibnamefont
  {Metzger}}, \bibinfo {author} {\bibfnamefont {D.}~\bibnamefont {Giannios}},\
  and\ \bibinfo {author} {\bibfnamefont {S.}~\bibnamefont {Horiuchi}},\
  }\bibfield  {title} {\bibinfo {title} {Heavy nuclei synthesized in gamma-ray
  burst outflows as the source of ultrahigh energy cosmic rays},\ }\href
  {https://doi.org/10.1111/j.1365-2966.2011.18873.x} {\bibfield  {journal}
  {\bibinfo  {journal} {Monthly Notices of the Royal Astronomical Society}\
  }\textbf {\bibinfo {volume} {415}},\ \bibinfo {pages} {2495} (\bibinfo {year}
  {2011}{\natexlab{a}})}\BibitemShut {NoStop}%
\bibitem [{\citenamefont {Horiuchi}\ \emph {et~al.}(2012)\citenamefont
  {Horiuchi}, \citenamefont {Murase}, \citenamefont {Ioka},\ and\ \citenamefont
  {Mészáros}}]{Horiuchi2012}%
  \BibitemOpen
  \bibfield  {author} {\bibinfo {author} {\bibfnamefont {S.}~\bibnamefont
  {Horiuchi}}, \bibinfo {author} {\bibfnamefont {K.}~\bibnamefont {Murase}},
  \bibinfo {author} {\bibfnamefont {K.}~\bibnamefont {Ioka}},\ and\ \bibinfo
  {author} {\bibfnamefont {P.}~\bibnamefont {Mészáros}},\ }\bibfield  {title}
  {\bibinfo {title} {The survival of nuclei in jets associated with
  core-collapse supernovae and gamma-ray bursts},\ }\href
  {https://doi.org/10.1088/0004-637x/753/1/69} {\bibfield  {journal} {\bibinfo
  {journal} {The Astrophysical Journal}\ }\textbf {\bibinfo {volume} {753}},\
  \bibinfo {pages} {69} (\bibinfo {year} {2012})}\BibitemShut {NoStop}%
\bibitem [{\citenamefont {Bhattacharya}\ \emph {et~al.}(2022)\citenamefont
  {Bhattacharya}, \citenamefont {Horiuchi},\ and\ \citenamefont
  {Murase}}]{MBh2021}%
  \BibitemOpen
  \bibfield  {author} {\bibinfo {author} {\bibfnamefont {M.}~\bibnamefont
  {Bhattacharya}}, \bibinfo {author} {\bibfnamefont {S.}~\bibnamefont
  {Horiuchi}},\ and\ \bibinfo {author} {\bibfnamefont {K.}~\bibnamefont
  {Murase}},\ }\bibfield  {title} {\bibinfo {title} {{On the synthesis of heavy
  nuclei in protomagnetar outflows and implications for ultra-high energy
  cosmic rays}},\ }\href {https://doi.org/10.1093/mnras/stac1721} {\bibfield
  {journal} {\bibinfo  {journal} {Mon. Not. Roy. Astron. Soc.}\ }\textbf
  {\bibinfo {volume} {514}},\ \bibinfo {pages} {6011} (\bibinfo {year}
  {2022})},\ \Eprint {https://arxiv.org/abs/2111.05863} {arXiv:2111.05863
  [astro-ph.HE]} \BibitemShut {NoStop}%
\bibitem [{\citenamefont {Ekanger}\ \emph {et~al.}(2022)\citenamefont
  {Ekanger}, \citenamefont {Bhattacharya},\ and\ \citenamefont
  {Horiuchi}}]{Ekanger:2022tia}%
  \BibitemOpen
  \bibfield  {author} {\bibinfo {author} {\bibfnamefont {N.}~\bibnamefont
  {Ekanger}}, \bibinfo {author} {\bibfnamefont {M.}~\bibnamefont
  {Bhattacharya}},\ and\ \bibinfo {author} {\bibfnamefont {S.}~\bibnamefont
  {Horiuchi}},\ }\bibfield  {title} {\bibinfo {title} {{Systematic exploration
  of heavy element nucleosynthesis in protomagnetar outflows}},\ }\href
  {https://doi.org/10.1093/mnras/stac896} {\bibfield  {journal} {\bibinfo
  {journal} {Mon. Not. Roy. Astron. Soc.}\ }\textbf {\bibinfo {volume} {513}},\
  \bibinfo {pages} {405} (\bibinfo {year} {2022})},\ \Eprint
  {https://arxiv.org/abs/2201.03576} {arXiv:2201.03576 [astro-ph.HE]}
  \BibitemShut {NoStop}%
\bibitem [{\citenamefont {Ekanger}\ \emph {et~al.}(2023)\citenamefont
  {Ekanger}, \citenamefont {Bhattacharya},\ and\ \citenamefont
  {Horiuchi}}]{Ekanger:2023mde}%
  \BibitemOpen
  \bibfield  {author} {\bibinfo {author} {\bibfnamefont {N.}~\bibnamefont
  {Ekanger}}, \bibinfo {author} {\bibfnamefont {M.}~\bibnamefont
  {Bhattacharya}},\ and\ \bibinfo {author} {\bibfnamefont {S.}~\bibnamefont
  {Horiuchi}},\ }\bibfield  {title} {\bibinfo {title} {{Nucleosynthesis in
  outflows of compact objects and detection prospects of associated
  kilonovae}},\ }\href {https://doi.org/10.1093/mnras/stad2348} {\bibfield
  {journal} {\bibinfo  {journal} {Mon. Not. Roy. Astron. Soc.}\ }\textbf
  {\bibinfo {volume} {525}},\ \bibinfo {pages} {2040} (\bibinfo {year}
  {2023})},\ \Eprint {https://arxiv.org/abs/2303.00765} {arXiv:2303.00765
  [astro-ph.HE]} \BibitemShut {NoStop}%
\bibitem [{\citenamefont {Thompson}\ \emph {et~al.}(2004)\citenamefont
  {Thompson}, \citenamefont {Chang},\ and\ \citenamefont
  {Quataert}}]{Thompson2004}%
  \BibitemOpen
  \bibfield  {author} {\bibinfo {author} {\bibfnamefont {T.~A.}\ \bibnamefont
  {Thompson}}, \bibinfo {author} {\bibfnamefont {P.}~\bibnamefont {Chang}},\
  and\ \bibinfo {author} {\bibfnamefont {E.}~\bibnamefont {Quataert}},\
  }\bibfield  {title} {\bibinfo {title} {Magnetar spin‐down, hyperenergetic
  supernovae, and gamma‐ray bursts},\ }\href {https://doi.org/10.1086/421969}
  {\bibfield  {journal} {\bibinfo  {journal} {The Astrophysical Journal}\
  }\textbf {\bibinfo {volume} {611}},\ \bibinfo {pages} {380} (\bibinfo {year}
  {2004})}\BibitemShut {NoStop}%
\bibitem [{\citenamefont {Wang}\ \emph {et~al.}(2008)\citenamefont {Wang},
  \citenamefont {Razzaque},\ and\ \citenamefont {Meszaros}}]{Wang:2007xj}%
  \BibitemOpen
  \bibfield  {author} {\bibinfo {author} {\bibfnamefont {X.-Y.}\ \bibnamefont
  {Wang}}, \bibinfo {author} {\bibfnamefont {S.}~\bibnamefont {Razzaque}},\
  and\ \bibinfo {author} {\bibfnamefont {P.}~\bibnamefont {Meszaros}},\
  }\bibfield  {title} {\bibinfo {title} {{On the Origin and Survival of UHE
  Cosmic-Ray Nuclei in GRBs and Hypernovae}},\ }\href
  {https://doi.org/10.1086/529018} {\bibfield  {journal} {\bibinfo  {journal}
  {Astrophys. J.}\ }\textbf {\bibinfo {volume} {677}},\ \bibinfo {pages} {432}
  (\bibinfo {year} {2008})},\ \Eprint {https://arxiv.org/abs/0711.2065}
  {arXiv:0711.2065 [astro-ph]} \BibitemShut {NoStop}%
\bibitem [{\citenamefont {{Puget}}\ \emph {et~al.}(1976)\citenamefont
  {{Puget}}, \citenamefont {{Stecker}},\ and\ \citenamefont
  {{Bredekamp}}}]{1976puget}%
  \BibitemOpen
  \bibfield  {author} {\bibinfo {author} {\bibfnamefont {J.~L.}\ \bibnamefont
  {{Puget}}}, \bibinfo {author} {\bibfnamefont {F.~W.}\ \bibnamefont
  {{Stecker}}},\ and\ \bibinfo {author} {\bibfnamefont {J.~H.}\ \bibnamefont
  {{Bredekamp}}},\ }\bibfield  {title} {\bibinfo {title} {{Photonuclear
  interactions of ultrahigh energy cosmic rays and their astrophysical
  consequences.}},\ }\href {https://doi.org/10.1086/154321} {\bibfield
  {journal} {\bibinfo  {journal} {\apj}\ }\textbf {\bibinfo {volume} {205}},\
  \bibinfo {pages} {638} (\bibinfo {year} {1976})}\BibitemShut {NoStop}%
\bibitem [{\citenamefont {Stecker}\ and\ \citenamefont
  {Salamon}(1999)}]{Stecker:1998ib}%
  \BibitemOpen
  \bibfield  {author} {\bibinfo {author} {\bibfnamefont {F.~W.}\ \bibnamefont
  {Stecker}}\ and\ \bibinfo {author} {\bibfnamefont {M.~H.}\ \bibnamefont
  {Salamon}},\ }\bibfield  {title} {\bibinfo {title} {{Photodisintegration of
  ultrahigh-energy cosmic rays: A New determination}},\ }\href
  {https://doi.org/10.1086/306816} {\bibfield  {journal} {\bibinfo  {journal}
  {Astrophys. J.}\ }\textbf {\bibinfo {volume} {512}},\ \bibinfo {pages} {521}
  (\bibinfo {year} {1999})},\ \Eprint {https://arxiv.org/abs/astro-ph/9808110}
  {arXiv:astro-ph/9808110} \BibitemShut {NoStop}%
\bibitem [{\citenamefont {Khan}\ \emph {et~al.}(2005)\citenamefont {Khan},
  \citenamefont {Goriely}, \citenamefont {Allard}, \citenamefont {Parizot},
  \citenamefont {Suomijarvi}, \citenamefont {Koning}, \citenamefont {Hilaire},\
  and\ \citenamefont {Duijvestijn}}]{Khan:2004nd}%
  \BibitemOpen
  \bibfield  {author} {\bibinfo {author} {\bibfnamefont {E.}~\bibnamefont
  {Khan}}, \bibinfo {author} {\bibfnamefont {S.}~\bibnamefont {Goriely}},
  \bibinfo {author} {\bibfnamefont {D.}~\bibnamefont {Allard}}, \bibinfo
  {author} {\bibfnamefont {E.}~\bibnamefont {Parizot}}, \bibinfo {author}
  {\bibfnamefont {T.}~\bibnamefont {Suomijarvi}}, \bibinfo {author}
  {\bibfnamefont {A.~J.}\ \bibnamefont {Koning}}, \bibinfo {author}
  {\bibfnamefont {S.}~\bibnamefont {Hilaire}},\ and\ \bibinfo {author}
  {\bibfnamefont {M.~C.}\ \bibnamefont {Duijvestijn}},\ }\bibfield  {title}
  {\bibinfo {title} {{Photodisintegration of ultra-high-energy cosmic rays
  revisited}},\ }\href {https://doi.org/10.1016/j.astropartphys.2004.12.007}
  {\bibfield  {journal} {\bibinfo  {journal} {Astropart. Phys.}\ }\textbf
  {\bibinfo {volume} {23}},\ \bibinfo {pages} {191} (\bibinfo {year} {2005})},\
  \Eprint {https://arxiv.org/abs/astro-ph/0412109} {arXiv:astro-ph/0412109}
  \BibitemShut {NoStop}%
\bibitem [{\citenamefont {Bhattacharya}\ \emph {et~al.}(2023)\citenamefont
  {Bhattacharya}, \citenamefont {Carpio}, \citenamefont {Murase},\ and\
  \citenamefont {Horiuchi}}]{Bhattacharya:2022btx}%
  \BibitemOpen
  \bibfield  {author} {\bibinfo {author} {\bibfnamefont {M.}~\bibnamefont
  {Bhattacharya}}, \bibinfo {author} {\bibfnamefont {J.~A.}\ \bibnamefont
  {Carpio}}, \bibinfo {author} {\bibfnamefont {K.}~\bibnamefont {Murase}},\
  and\ \bibinfo {author} {\bibfnamefont {S.}~\bibnamefont {Horiuchi}},\
  }\bibfield  {title} {\bibinfo {title} {{High-energy neutrino emission from
  magnetised jets of rapidly rotating protomagnetars}},\ }\href
  {https://doi.org/10.1093/mnras/stad494} {\bibfield  {journal} {\bibinfo
  {journal} {Mon. Not. Roy. Astron. Soc.}\ }\textbf {\bibinfo {volume} {521}},\
  \bibinfo {pages} {2391} (\bibinfo {year} {2023})},\ \Eprint
  {https://arxiv.org/abs/2210.08029} {arXiv:2210.08029 [astro-ph.HE]}
  \BibitemShut {NoStop}%
\bibitem [{\citenamefont {{Soderberg}}\ \emph {et~al.}(2006)\citenamefont
  {{Soderberg}}, \citenamefont {{Kulkarni}}, \citenamefont {{Nakar}},
  \citenamefont {{Berger}}, \citenamefont {{Cameron}}, \citenamefont {{Fox}},
  \citenamefont {{Frail}}, \citenamefont {{Gal-Yam}}, \citenamefont {{Sari}},
  \citenamefont {{Cenko}}, \citenamefont {{Kasliwal}}, \citenamefont
  {{Chevalier}}, \citenamefont {{Piran}}, \citenamefont {{Price}},
  \citenamefont {{Schmidt}}, \citenamefont {{Pooley}}, \citenamefont {{Moon}},
  \citenamefont {{Penprase}}, \citenamefont {{Ofek}}, \citenamefont {{Rau}},
  \citenamefont {{Gehrels}}, \citenamefont {{Nousek}}, \citenamefont
  {{Burrows}}, \citenamefont {{Persson}},\ and\ \citenamefont
  {{McCarthy}}}]{Soderberg2006}%
  \BibitemOpen
  \bibfield  {author} {\bibinfo {author} {\bibfnamefont {A.~M.}\ \bibnamefont
  {{Soderberg}}}, \bibinfo {author} {\bibfnamefont {S.~R.}\ \bibnamefont
  {{Kulkarni}}}, \bibinfo {author} {\bibfnamefont {E.}~\bibnamefont {{Nakar}}},
  \bibinfo {author} {\bibfnamefont {E.}~\bibnamefont {{Berger}}}, \bibinfo
  {author} {\bibfnamefont {P.~B.}\ \bibnamefont {{Cameron}}}, \bibinfo {author}
  {\bibfnamefont {D.~B.}\ \bibnamefont {{Fox}}}, \bibinfo {author}
  {\bibfnamefont {D.}~\bibnamefont {{Frail}}}, \bibinfo {author} {\bibfnamefont
  {A.}~\bibnamefont {{Gal-Yam}}}, \bibinfo {author} {\bibfnamefont
  {R.}~\bibnamefont {{Sari}}}, \bibinfo {author} {\bibfnamefont {S.~B.}\
  \bibnamefont {{Cenko}}}, \bibinfo {author} {\bibfnamefont {M.}~\bibnamefont
  {{Kasliwal}}}, \bibinfo {author} {\bibfnamefont {R.~A.}\ \bibnamefont
  {{Chevalier}}}, \bibinfo {author} {\bibfnamefont {T.}~\bibnamefont
  {{Piran}}}, \bibinfo {author} {\bibfnamefont {P.~A.}\ \bibnamefont
  {{Price}}}, \bibinfo {author} {\bibfnamefont {B.~P.}\ \bibnamefont
  {{Schmidt}}}, \bibinfo {author} {\bibfnamefont {G.}~\bibnamefont {{Pooley}}},
  \bibinfo {author} {\bibfnamefont {D.~S.}\ \bibnamefont {{Moon}}}, \bibinfo
  {author} {\bibfnamefont {B.~E.}\ \bibnamefont {{Penprase}}}, \bibinfo
  {author} {\bibfnamefont {E.}~\bibnamefont {{Ofek}}}, \bibinfo {author}
  {\bibfnamefont {A.}~\bibnamefont {{Rau}}}, \bibinfo {author} {\bibfnamefont
  {N.}~\bibnamefont {{Gehrels}}}, \bibinfo {author} {\bibfnamefont {J.~A.}\
  \bibnamefont {{Nousek}}}, \bibinfo {author} {\bibfnamefont {D.~N.}\
  \bibnamefont {{Burrows}}}, \bibinfo {author} {\bibfnamefont {S.~E.}\
  \bibnamefont {{Persson}}},\ and\ \bibinfo {author} {\bibfnamefont {P.~J.}\
  \bibnamefont {{McCarthy}}},\ }\bibfield  {title} {\bibinfo {title}
  {{Relativistic ejecta from X-ray flash XRF 060218 and the rate of cosmic
  explosions}},\ }\href {https://doi.org/10.1038/nature05087} {\bibfield
  {journal} {\bibinfo  {journal} {\nat}\ }\textbf {\bibinfo {volume} {442}},\
  \bibinfo {pages} {1014} (\bibinfo {year} {2006})},\ \Eprint
  {https://arxiv.org/abs/astro-ph/0604389} {arXiv:astro-ph/0604389 [astro-ph]}
  \BibitemShut {NoStop}%
\bibitem [{\citenamefont {{Margutti}}\ \emph {et~al.}(2013)\citenamefont
  {{Margutti}}, \citenamefont {{Soderberg}}, \citenamefont {{Wieringa}},
  \citenamefont {{Edwards}}, \citenamefont {{Chevalier}}, \citenamefont
  {{Morsony}}, \citenamefont {{Barniol Duran}}, \citenamefont {{Sironi}},
  \citenamefont {{Zauderer}}, \citenamefont {{Milisavljevic}}, \citenamefont
  {{Kamble}},\ and\ \citenamefont {{Pian}}}]{Margutti2013}%
  \BibitemOpen
  \bibfield  {author} {\bibinfo {author} {\bibfnamefont {R.}~\bibnamefont
  {{Margutti}}}, \bibinfo {author} {\bibfnamefont {A.~M.}\ \bibnamefont
  {{Soderberg}}}, \bibinfo {author} {\bibfnamefont {M.~H.}\ \bibnamefont
  {{Wieringa}}}, \bibinfo {author} {\bibfnamefont {P.~G.}\ \bibnamefont
  {{Edwards}}}, \bibinfo {author} {\bibfnamefont {R.~A.}\ \bibnamefont
  {{Chevalier}}}, \bibinfo {author} {\bibfnamefont {B.~J.}\ \bibnamefont
  {{Morsony}}}, \bibinfo {author} {\bibfnamefont {R.}~\bibnamefont {{Barniol
  Duran}}}, \bibinfo {author} {\bibfnamefont {L.}~\bibnamefont {{Sironi}}},
  \bibinfo {author} {\bibfnamefont {B.~A.}\ \bibnamefont {{Zauderer}}},
  \bibinfo {author} {\bibfnamefont {D.}~\bibnamefont {{Milisavljevic}}},
  \bibinfo {author} {\bibfnamefont {A.}~\bibnamefont {{Kamble}}},\ and\
  \bibinfo {author} {\bibfnamefont {E.}~\bibnamefont {{Pian}}},\ }\bibfield
  {title} {\bibinfo {title} {{The Signature of the Central Engine in the
  Weakest Relativistic Explosions: GRB 100316D}},\ }\href
  {https://doi.org/10.1088/0004-637X/778/1/18} {\bibfield  {journal} {\bibinfo
  {journal} {\apj}\ }\textbf {\bibinfo {volume} {778}},\ \bibinfo {eid} {18}
  (\bibinfo {year} {2013})},\ \Eprint {https://arxiv.org/abs/1308.1687}
  {arXiv:1308.1687 [astro-ph.HE]} \BibitemShut {NoStop}%
\bibitem [{\citenamefont {{Margutti}}\ \emph {et~al.}(2014)\citenamefont
  {{Margutti}}, \citenamefont {{Milisavljevic}}, \citenamefont {{Soderberg}},
  \citenamefont {{Guidorzi}}, \citenamefont {{Morsony}}, \citenamefont
  {{Sanders}}, \citenamefont {{Chakraborti}}, \citenamefont {{Ray}},
  \citenamefont {{Kamble}}, \citenamefont {{Drout}}, \citenamefont {{Parrent}},
  \citenamefont {{Zauderer}},\ and\ \citenamefont {{Chomiuk}}}]{Margutti2014}%
  \BibitemOpen
  \bibfield  {author} {\bibinfo {author} {\bibfnamefont {R.}~\bibnamefont
  {{Margutti}}}, \bibinfo {author} {\bibfnamefont {D.}~\bibnamefont
  {{Milisavljevic}}}, \bibinfo {author} {\bibfnamefont {A.~M.}\ \bibnamefont
  {{Soderberg}}}, \bibinfo {author} {\bibfnamefont {C.}~\bibnamefont
  {{Guidorzi}}}, \bibinfo {author} {\bibfnamefont {B.~J.}\ \bibnamefont
  {{Morsony}}}, \bibinfo {author} {\bibfnamefont {N.}~\bibnamefont
  {{Sanders}}}, \bibinfo {author} {\bibfnamefont {S.}~\bibnamefont
  {{Chakraborti}}}, \bibinfo {author} {\bibfnamefont {A.}~\bibnamefont
  {{Ray}}}, \bibinfo {author} {\bibfnamefont {A.}~\bibnamefont {{Kamble}}},
  \bibinfo {author} {\bibfnamefont {M.}~\bibnamefont {{Drout}}}, \bibinfo
  {author} {\bibfnamefont {J.}~\bibnamefont {{Parrent}}}, \bibinfo {author}
  {\bibfnamefont {A.}~\bibnamefont {{Zauderer}}},\ and\ \bibinfo {author}
  {\bibfnamefont {L.}~\bibnamefont {{Chomiuk}}},\ }\bibfield  {title} {\bibinfo
  {title} {{Relativistic Supernovae have Shorter-lived Central Engines or More
  Extended Progenitors: The Case of SN 2012ap}},\ }\href
  {https://doi.org/10.1088/0004-637X/797/2/107} {\bibfield  {journal} {\bibinfo
   {journal} {\apj}\ }\textbf {\bibinfo {volume} {797}},\ \bibinfo {eid} {107}
  (\bibinfo {year} {2014})},\ \Eprint {https://arxiv.org/abs/1402.6344}
  {arXiv:1402.6344 [astro-ph.HE]} \BibitemShut {NoStop}%
\bibitem [{\citenamefont {Murase}\ \emph {et~al.}(2014)\citenamefont {Murase},
  \citenamefont {Dasgupta},\ and\ \citenamefont {Thompson}}]{Murase:2013mpa}%
  \BibitemOpen
  \bibfield  {author} {\bibinfo {author} {\bibfnamefont {K.}~\bibnamefont
  {Murase}}, \bibinfo {author} {\bibfnamefont {B.}~\bibnamefont {Dasgupta}},\
  and\ \bibinfo {author} {\bibfnamefont {T.~A.}\ \bibnamefont {Thompson}},\
  }\bibfield  {title} {\bibinfo {title} {{Quasithermal Neutrinos from Rotating
  Protoneutron Stars Born during Core Collapse of Massive Stars}},\ }\href
  {https://doi.org/10.1103/PhysRevD.89.043012} {\bibfield  {journal} {\bibinfo
  {journal} {Phys. Rev. D}\ }\textbf {\bibinfo {volume} {89}},\ \bibinfo
  {pages} {043012} (\bibinfo {year} {2014})},\ \Eprint
  {https://arxiv.org/abs/1303.2612} {arXiv:1303.2612 [astro-ph.HE]}
  \BibitemShut {NoStop}%
\bibitem [{\citenamefont {Carpio}\ \emph {et~al.}(2024)\citenamefont {Carpio},
  \citenamefont {Ekanger}, \citenamefont {Bhattacharya}, \citenamefont
  {Murase},\ and\ \citenamefont {Horiuchi}}]{Carpio:2023wyt}%
  \BibitemOpen
  \bibfield  {author} {\bibinfo {author} {\bibfnamefont {J.~A.}\ \bibnamefont
  {Carpio}}, \bibinfo {author} {\bibfnamefont {N.}~\bibnamefont {Ekanger}},
  \bibinfo {author} {\bibfnamefont {M.}~\bibnamefont {Bhattacharya}}, \bibinfo
  {author} {\bibfnamefont {K.}~\bibnamefont {Murase}},\ and\ \bibinfo {author}
  {\bibfnamefont {S.}~\bibnamefont {Horiuchi}},\ }\bibfield  {title} {\bibinfo
  {title} {{Quasithermal GeV neutrinos from neutron-loaded magnetized outflows
  in core-collapse supernovae: Spectra and light curves}},\ }\href
  {https://doi.org/10.1103/PhysRevD.110.083012} {\bibfield  {journal} {\bibinfo
   {journal} {Phys. Rev. D}\ }\textbf {\bibinfo {volume} {110}},\ \bibinfo
  {pages} {083012} (\bibinfo {year} {2024})},\ \Eprint
  {https://arxiv.org/abs/2310.16823} {arXiv:2310.16823 [astro-ph.HE]}
  \BibitemShut {NoStop}%
\bibitem [{\citenamefont {{Murase}}\ and\ \citenamefont
  {{Beacom}}(2010)}]{MuraseBeacom2010}%
  \BibitemOpen
  \bibfield  {author} {\bibinfo {author} {\bibfnamefont {K.}~\bibnamefont
  {{Murase}}}\ and\ \bibinfo {author} {\bibfnamefont {J.~F.}\ \bibnamefont
  {{Beacom}}},\ }\bibfield  {title} {\bibinfo {title} {{Neutrino background
  flux from sources of ultrahigh-energy cosmic-ray nuclei}},\ }\href
  {https://doi.org/10.1103/PhysRevD.81.123001} {\bibfield  {journal} {\bibinfo
  {journal} {\prd}\ }\textbf {\bibinfo {volume} {81}},\ \bibinfo {eid} {123001}
  (\bibinfo {year} {2010})},\ \Eprint {https://arxiv.org/abs/1003.4959}
  {arXiv:1003.4959 [astro-ph.HE]} \BibitemShut {NoStop}%
\bibitem [{\citenamefont {Murase}\ and\ \citenamefont
  {Beacom}(2010)}]{Murase:2010va}%
  \BibitemOpen
  \bibfield  {author} {\bibinfo {author} {\bibfnamefont {K.}~\bibnamefont
  {Murase}}\ and\ \bibinfo {author} {\bibfnamefont {J.~F.}\ \bibnamefont
  {Beacom}},\ }\bibfield  {title} {\bibinfo {title} {{Very-High-Energy
  Gamma-Ray Signal from Nuclear Photodisintegration as a Probe of Extragalactic
  Sources of Ultrahigh-Energy Nuclei}},\ }\href
  {https://doi.org/10.1103/PhysRevD.82.043008} {\bibfield  {journal} {\bibinfo
  {journal} {Phys. Rev. D}\ }\textbf {\bibinfo {volume} {82}},\ \bibinfo
  {pages} {043008} (\bibinfo {year} {2010})},\ \Eprint
  {https://arxiv.org/abs/1002.3980} {arXiv:1002.3980 [astro-ph.HE]}
  \BibitemShut {NoStop}%
\bibitem [{\citenamefont {Zhang}\ and\ \citenamefont
  {Murase}(2023)}]{Zhang:2023ewt}%
  \BibitemOpen
  \bibfield  {author} {\bibinfo {author} {\bibfnamefont {B.~T.}\ \bibnamefont
  {Zhang}}\ and\ \bibinfo {author} {\bibfnamefont {K.}~\bibnamefont {Murase}},\
  }\bibfield  {title} {\bibinfo {title} {{Nuclear and electromagnetic cascades
  induced by ultra-high-energy cosmic rays in radio galaxies: implications for
  Centaurus A}},\ }\href {https://doi.org/10.1093/mnras/stad1829} {\bibfield
  {journal} {\bibinfo  {journal} {Mon. Not. Roy. Astron. Soc.}\ }\textbf
  {\bibinfo {volume} {524}},\ \bibinfo {pages} {76} (\bibinfo {year} {2023})},\
  \Eprint {https://arxiv.org/abs/2302.14048} {arXiv:2302.14048 [astro-ph.HE]}
  \BibitemShut {NoStop}%
\bibitem [{\citenamefont {Karakula}\ and\ \citenamefont
  {Tkaczyk}(1993)}]{KARAKULA1993229}%
  \BibitemOpen
  \bibfield  {author} {\bibinfo {author} {\bibfnamefont {S.}~\bibnamefont
  {Karakula}}\ and\ \bibinfo {author} {\bibfnamefont {W.}~\bibnamefont
  {Tkaczyk}},\ }\bibfield  {title} {\bibinfo {title} {The formation of the
  cosmic ray energy spectrum by a photon field},\ }\href
  {https://doi.org/https://doi.org/10.1016/0927-6505(93)90023-7} {\bibfield
  {journal} {\bibinfo  {journal} {Astroparticle Physics}\ }\textbf {\bibinfo
  {volume} {1}},\ \bibinfo {pages} {229} (\bibinfo {year} {1993})}\BibitemShut
  {NoStop}%
\bibitem [{\citenamefont {{Koning}}\ and\ \citenamefont
  {{Rochman}}(2012)}]{2012NDS...113.2841K}%
  \BibitemOpen
  \bibfield  {author} {\bibinfo {author} {\bibfnamefont {A.~J.}\ \bibnamefont
  {{Koning}}}\ and\ \bibinfo {author} {\bibfnamefont {D.}~\bibnamefont
  {{Rochman}}},\ }\bibfield  {title} {\bibinfo {title} {{Modern Nuclear Data
  Evaluation with the TALYS Code System}},\ }\href
  {https://doi.org/10.1016/j.nds.2012.11.002} {\bibfield  {journal} {\bibinfo
  {journal} {Nuclear Data Sheets}\ }\textbf {\bibinfo {volume} {113}},\
  \bibinfo {pages} {2841} (\bibinfo {year} {2012})}\BibitemShut {NoStop}%
\bibitem [{\citenamefont {{Koning}}\ \emph {et~al.}(2019)\citenamefont
  {{Koning}}, \citenamefont {{Rochman}}, \citenamefont {{Sublet}},
  \citenamefont {{Dzysiuk}}, \citenamefont {{Fleming}},\ and\ \citenamefont
  {{van der Marck}}}]{2019NDS...155....1K}%
  \BibitemOpen
  \bibfield  {author} {\bibinfo {author} {\bibfnamefont {A.~J.}\ \bibnamefont
  {{Koning}}}, \bibinfo {author} {\bibfnamefont {D.}~\bibnamefont {{Rochman}}},
  \bibinfo {author} {\bibfnamefont {J.~C.}\ \bibnamefont {{Sublet}}}, \bibinfo
  {author} {\bibfnamefont {N.}~\bibnamefont {{Dzysiuk}}}, \bibinfo {author}
  {\bibfnamefont {M.}~\bibnamefont {{Fleming}}},\ and\ \bibinfo {author}
  {\bibfnamefont {S.}~\bibnamefont {{van der Marck}}},\ }\bibfield  {title}
  {\bibinfo {title} {{TENDL: Complete Nuclear Data Library for Innovative
  Nuclear Science and Technology}},\ }\href
  {https://doi.org/10.1016/j.nds.2019.01.002} {\bibfield  {journal} {\bibinfo
  {journal} {Nuclear Data Sheets}\ }\textbf {\bibinfo {volume} {155}},\
  \bibinfo {pages} {1} (\bibinfo {year} {2019})}\BibitemShut {NoStop}%
\bibitem [{\citenamefont {Firestone}(2020)}]{Firestone:2020rty}%
  \BibitemOpen
  \bibfield  {author} {\bibinfo {author} {\bibfnamefont {R.~B.}\ \bibnamefont
  {Firestone}},\ }\href@noop {} {\bibinfo {title} {{The Origin of the Giant
  Dipole Resonance}}} (\bibinfo {year} {2020}),\ \Eprint
  {https://arxiv.org/abs/2009.03356} {arXiv:2009.03356 [nucl-th]} \BibitemShut
  {NoStop}%
\bibitem [{\citenamefont {Goldhaber}\ and\ \citenamefont
  {Teller}(1948)}]{goldhaberteller1948}%
  \BibitemOpen
  \bibfield  {author} {\bibinfo {author} {\bibfnamefont {M.}~\bibnamefont
  {Goldhaber}}\ and\ \bibinfo {author} {\bibfnamefont {E.}~\bibnamefont
  {Teller}},\ }\bibfield  {title} {\bibinfo {title} {On nuclear dipole
  vibrations},\ }\href {https://doi.org/10.1103/PhysRev.74.1046} {\bibfield
  {journal} {\bibinfo  {journal} {Phys. Rev.}\ }\textbf {\bibinfo {volume}
  {74}},\ \bibinfo {pages} {1046} (\bibinfo {year} {1948})}\BibitemShut
  {NoStop}%
\bibitem [{\citenamefont {Agostinelli}\ \emph {et~al.}(2003)\citenamefont
  {Agostinelli} \emph {et~al.}}]{GEANT4:2002zbu}%
  \BibitemOpen
  \bibfield  {author} {\bibinfo {author} {\bibfnamefont {S.}~\bibnamefont
  {Agostinelli}} \emph {et~al.} (\bibinfo {collaboration} {GEANT4}),\
  }\bibfield  {title} {\bibinfo {title} {{GEANT4--a simulation toolkit}},\
  }\href {https://doi.org/10.1016/S0168-9002(03)01368-8} {\bibfield  {journal}
  {\bibinfo  {journal} {Nucl. Instrum. Meth. A}\ }\textbf {\bibinfo {volume}
  {506}},\ \bibinfo {pages} {250} (\bibinfo {year} {2003})}\BibitemShut
  {NoStop}%
\bibitem [{\citenamefont {{Rachen}}(1996)}]{1996PhDT........59R}%
  \BibitemOpen
  \bibfield  {author} {\bibinfo {author} {\bibfnamefont {J.~P.}\ \bibnamefont
  {{Rachen}}},\ }\emph {\bibinfo {title} {{Interaction Processes and
  Statistical Properties of the Propagation of Cosmic Rays in Photon
  Backgrounds}}},\ \href@noop {} {Ph.D. thesis},\ \bibinfo  {school}
  {Max-Planck-Institute for Radioastronomy, Bonn} (\bibinfo {year}
  {1996})\BibitemShut {NoStop}%
\bibitem [{\citenamefont {Allard}\ \emph {et~al.}(2005)\citenamefont {Allard},
  \citenamefont {Parizot}, \citenamefont {Khan}, \citenamefont {Goriely},\ and\
  \citenamefont {Olinto}}]{Allard:2005ha}%
  \BibitemOpen
  \bibfield  {author} {\bibinfo {author} {\bibfnamefont {D.}~\bibnamefont
  {Allard}}, \bibinfo {author} {\bibfnamefont {E.}~\bibnamefont {Parizot}},
  \bibinfo {author} {\bibfnamefont {E.}~\bibnamefont {Khan}}, \bibinfo {author}
  {\bibfnamefont {S.}~\bibnamefont {Goriely}},\ and\ \bibinfo {author}
  {\bibfnamefont {A.~V.}\ \bibnamefont {Olinto}},\ }\bibfield  {title}
  {\bibinfo {title} {{UHE nuclei propagation and the interpretation of the
  ankle in the cosmic-ray spectrum}},\ }\href
  {https://doi.org/10.1051/0004-6361:200500199} {\bibfield  {journal} {\bibinfo
   {journal} {Astron. Astrophys.}\ }\textbf {\bibinfo {volume} {443}},\
  \bibinfo {pages} {L29} (\bibinfo {year} {2005})},\ \Eprint
  {https://arxiv.org/abs/astro-ph/0505566} {arXiv:astro-ph/0505566}
  \BibitemShut {NoStop}%
\bibitem [{\citenamefont {{Goldreich}}\ and\ \citenamefont
  {{Julian}}(1969)}]{1969ApJ...157..869G}%
  \BibitemOpen
  \bibfield  {author} {\bibinfo {author} {\bibfnamefont {P.}~\bibnamefont
  {{Goldreich}}}\ and\ \bibinfo {author} {\bibfnamefont {W.~H.}\ \bibnamefont
  {{Julian}}},\ }\bibfield  {title} {\bibinfo {title} {{Pulsar
  Electrodynamics}},\ }\href {https://doi.org/10.1086/150119} {\bibfield
  {journal} {\bibinfo  {journal} {\apj}\ }\textbf {\bibinfo {volume} {157}},\
  \bibinfo {pages} {869} (\bibinfo {year} {1969})}\BibitemShut {NoStop}%
\bibitem [{\citenamefont {{Bucciantini}}\ \emph {et~al.}(2011)\citenamefont
  {{Bucciantini}}, \citenamefont {{Arons}},\ and\ \citenamefont
  {{Amato}}}]{bucciantini11}%
  \BibitemOpen
  \bibfield  {author} {\bibinfo {author} {\bibfnamefont {N.}~\bibnamefont
  {{Bucciantini}}}, \bibinfo {author} {\bibfnamefont {J.}~\bibnamefont
  {{Arons}}},\ and\ \bibinfo {author} {\bibfnamefont {E.}~\bibnamefont
  {{Amato}}},\ }\bibfield  {title} {\bibinfo {title} {{Modelling spectral
  evolution of pulsar wind nebulae inside supernova remnants}},\ }\href
  {https://doi.org/10.1111/j.1365-2966.2010.17449.x} {\bibfield  {journal}
  {\bibinfo  {journal} {\mnras}\ }\textbf {\bibinfo {volume} {410}},\ \bibinfo
  {pages} {381} (\bibinfo {year} {2011})},\ \Eprint
  {https://arxiv.org/abs/1005.1831} {arXiv:1005.1831 [astro-ph.HE]}
  \BibitemShut {NoStop}%
\bibitem [{\citenamefont {Timokhin}\ and\ \citenamefont
  {Harding}(2015)}]{Timokhin:2015dua}%
  \BibitemOpen
  \bibfield  {author} {\bibinfo {author} {\bibfnamefont {A.~N.}\ \bibnamefont
  {Timokhin}}\ and\ \bibinfo {author} {\bibfnamefont {A.~K.}\ \bibnamefont
  {Harding}},\ }\bibfield  {title} {\bibinfo {title} {{On the polar cap cascade
  pair multiplicity of young pulsars}},\ }\href
  {https://doi.org/10.1088/0004-637X/810/2/144} {\bibfield  {journal} {\bibinfo
   {journal} {Astrophys. J.}\ }\textbf {\bibinfo {volume} {810}},\ \bibinfo
  {pages} {144} (\bibinfo {year} {2015})},\ \Eprint
  {https://arxiv.org/abs/1504.02194} {arXiv:1504.02194 [astro-ph.HE]}
  \BibitemShut {NoStop}%
\bibitem [{\citenamefont {Murase}\ \emph {et~al.}(2015)\citenamefont {Murase},
  \citenamefont {Kashiyama}, \citenamefont {Kiuchi},\ and\ \citenamefont
  {Bartos}}]{Murase:2014bfa}%
  \BibitemOpen
  \bibfield  {author} {\bibinfo {author} {\bibfnamefont {K.}~\bibnamefont
  {Murase}}, \bibinfo {author} {\bibfnamefont {K.}~\bibnamefont {Kashiyama}},
  \bibinfo {author} {\bibfnamefont {K.}~\bibnamefont {Kiuchi}},\ and\ \bibinfo
  {author} {\bibfnamefont {I.}~\bibnamefont {Bartos}},\ }\bibfield  {title}
  {\bibinfo {title} {{Gamma-Ray and Hard X-Ray Emission from Pulsar-Aided
  Supernovae as a Probe of Particle Acceleration in Embryonic Pulsar Wind
  Nebulae}},\ }\href {https://doi.org/10.1088/0004-637X/805/1/82} {\bibfield
  {journal} {\bibinfo  {journal} {Astrophys. J.}\ }\textbf {\bibinfo {volume}
  {805}},\ \bibinfo {pages} {82} (\bibinfo {year} {2015})},\ \Eprint
  {https://arxiv.org/abs/1411.0619} {arXiv:1411.0619 [astro-ph.HE]}
  \BibitemShut {NoStop}%
\bibitem [{\citenamefont {{Kashiyama}}\ \emph {et~al.}(2016)\citenamefont
  {{Kashiyama}}, \citenamefont {{Murase}}, \citenamefont {{Bartos}},
  \citenamefont {{Kiuchi}},\ and\ \citenamefont {{Margutti}}}]{Kashiyama2016}%
  \BibitemOpen
  \bibfield  {author} {\bibinfo {author} {\bibfnamefont {K.}~\bibnamefont
  {{Kashiyama}}}, \bibinfo {author} {\bibfnamefont {K.}~\bibnamefont
  {{Murase}}}, \bibinfo {author} {\bibfnamefont {I.}~\bibnamefont {{Bartos}}},
  \bibinfo {author} {\bibfnamefont {K.}~\bibnamefont {{Kiuchi}}},\ and\
  \bibinfo {author} {\bibfnamefont {R.}~\bibnamefont {{Margutti}}},\ }\bibfield
   {title} {\bibinfo {title} {{Multi-messenger Tests for Fast-spinning Newborn
  Pulsars Embedded in Stripped-envelope Supernovae}},\ }\href
  {https://doi.org/10.3847/0004-637X/818/1/94} {\bibfield  {journal} {\bibinfo
  {journal} {\apj}\ }\textbf {\bibinfo {volume} {818}},\ \bibinfo {eid} {94}
  (\bibinfo {year} {2016})},\ \Eprint {https://arxiv.org/abs/1508.04393}
  {arXiv:1508.04393 [astro-ph.HE]} \BibitemShut {NoStop}%
\bibitem [{\citenamefont {Metzger}\ \emph
  {et~al.}(2011{\natexlab{b}})\citenamefont {Metzger}, \citenamefont
  {Giannios}, \citenamefont {Thompson}, \citenamefont {Bucciantini},\ and\
  \citenamefont {Quataert}}]{Metzger2011a}%
  \BibitemOpen
  \bibfield  {author} {\bibinfo {author} {\bibfnamefont {B.~D.}\ \bibnamefont
  {Metzger}}, \bibinfo {author} {\bibfnamefont {D.}~\bibnamefont {Giannios}},
  \bibinfo {author} {\bibfnamefont {T.~A.}\ \bibnamefont {Thompson}}, \bibinfo
  {author} {\bibfnamefont {N.}~\bibnamefont {Bucciantini}},\ and\ \bibinfo
  {author} {\bibfnamefont {E.}~\bibnamefont {Quataert}},\ }\bibfield  {title}
  {\bibinfo {title} {The protomagnetar model for gamma-ray bursts},\ }\href
  {https://doi.org/10.1111/j.1365-2966.2011.18280.x} {\bibfield  {journal}
  {\bibinfo  {journal} {Monthly Notices of the Royal Astronomical Society}\
  }\textbf {\bibinfo {volume} {413}},\ \bibinfo {pages} {2031} (\bibinfo {year}
  {2011}{\natexlab{b}})}\BibitemShut {NoStop}%
\bibitem [{\citenamefont {Pons}\ \emph {et~al.}(1999)\citenamefont {Pons},
  \citenamefont {Reddy}, \citenamefont {Prakash}, \citenamefont {Lattimer},\
  and\ \citenamefont {Miralles}}]{Pons1999}%
  \BibitemOpen
  \bibfield  {author} {\bibinfo {author} {\bibfnamefont {J.~A.}\ \bibnamefont
  {Pons}}, \bibinfo {author} {\bibfnamefont {S.}~\bibnamefont {Reddy}},
  \bibinfo {author} {\bibfnamefont {M.}~\bibnamefont {Prakash}}, \bibinfo
  {author} {\bibfnamefont {J.~M.}\ \bibnamefont {Lattimer}},\ and\ \bibinfo
  {author} {\bibfnamefont {J.~A.}\ \bibnamefont {Miralles}},\ }\bibfield
  {title} {\bibinfo {title} {Evolution of proto–neutron stars},\ }\href
  {https://doi.org/10.1086/306889} {\bibfield  {journal} {\bibinfo  {journal}
  {The Astrophysical Journal}\ }\textbf {\bibinfo {volume} {513}},\ \bibinfo
  {pages} {780} (\bibinfo {year} {1999})}\BibitemShut {NoStop}%
\bibitem [{\citenamefont {{Drenkhahn}}(2002)}]{Drenkhahn2002}%
  \BibitemOpen
  \bibfield  {author} {\bibinfo {author} {\bibfnamefont {G.}~\bibnamefont
  {{Drenkhahn}}},\ }\bibfield  {title} {\bibinfo {title} {{Acceleration of GRB
  outflows by Poynting flux dissipation}},\ }\href
  {https://doi.org/10.1051/0004-6361:20020390} {\bibfield  {journal} {\bibinfo
  {journal} {\aap}\ }\textbf {\bibinfo {volume} {387}},\ \bibinfo {pages} {714}
  (\bibinfo {year} {2002})},\ \Eprint {https://arxiv.org/abs/astro-ph/0112509}
  {arXiv:astro-ph/0112509 [astro-ph]} \BibitemShut {NoStop}%
\bibitem [{\citenamefont {Bromberg}\ \emph {et~al.}(2015)\citenamefont
  {Bromberg}, \citenamefont {Granot},\ and\ \citenamefont
  {Piran}}]{Bromberg:2014sja}%
  \BibitemOpen
  \bibfield  {author} {\bibinfo {author} {\bibfnamefont {O.}~\bibnamefont
  {Bromberg}}, \bibinfo {author} {\bibfnamefont {J.}~\bibnamefont {Granot}},\
  and\ \bibinfo {author} {\bibfnamefont {T.}~\bibnamefont {Piran}},\ }\bibfield
   {title} {\bibinfo {title} {{On the composition of GRBs{\textquoteright}
  Collapsar jets}},\ }\href {https://doi.org/10.1093/mnras/stv226} {\bibfield
  {journal} {\bibinfo  {journal} {Mon. Not. Roy. Astron. Soc.}\ }\textbf
  {\bibinfo {volume} {450}},\ \bibinfo {pages} {1077} (\bibinfo {year}
  {2015})},\ \Eprint {https://arxiv.org/abs/1407.0123} {arXiv:1407.0123
  [astro-ph.HE]} \BibitemShut {NoStop}%
\bibitem [{\citenamefont {Murase}\ \emph {et~al.}(2022)\citenamefont {Murase},
  \citenamefont {Mukhopadhyay}, \citenamefont {Kheirandish}, \citenamefont
  {Kimura},\ and\ \citenamefont {Fang}}]{Murase:2022vqf}%
  \BibitemOpen
  \bibfield  {author} {\bibinfo {author} {\bibfnamefont {K.}~\bibnamefont
  {Murase}}, \bibinfo {author} {\bibfnamefont {M.}~\bibnamefont
  {Mukhopadhyay}}, \bibinfo {author} {\bibfnamefont {A.}~\bibnamefont
  {Kheirandish}}, \bibinfo {author} {\bibfnamefont {S.~S.}\ \bibnamefont
  {Kimura}},\ and\ \bibinfo {author} {\bibfnamefont {K.}~\bibnamefont {Fang}},\
  }\bibfield  {title} {\bibinfo {title} {{Neutrinos from the Brightest
  Gamma-Ray Burst?}},\ }\href {https://doi.org/10.3847/2041-8213/aca3ae}
  {\bibfield  {journal} {\bibinfo  {journal} {Astrophys. J. Lett.}\ }\textbf
  {\bibinfo {volume} {941}},\ \bibinfo {pages} {L10} (\bibinfo {year}
  {2022})},\ \Eprint {https://arxiv.org/abs/2210.15625} {arXiv:2210.15625
  [astro-ph.HE]} \BibitemShut {NoStop}%
\bibitem [{\citenamefont {Giannios}\ and\ \citenamefont
  {Spruit}(2007)}]{Giannios:2006jb}%
  \BibitemOpen
  \bibfield  {author} {\bibinfo {author} {\bibfnamefont {D.}~\bibnamefont
  {Giannios}}\ and\ \bibinfo {author} {\bibfnamefont {H.~C.}\ \bibnamefont
  {Spruit}},\ }\bibfield  {title} {\bibinfo {title} {{Spectral and timing
  properties of a dissipative GRB photosphere}},\ }\href
  {https://doi.org/10.1051/0004-6361:20066739} {\bibfield  {journal} {\bibinfo
  {journal} {Astron. Astrophys.}\ }\textbf {\bibinfo {volume} {469}},\ \bibinfo
  {pages} {1} (\bibinfo {year} {2007})},\ \Eprint
  {https://arxiv.org/abs/astro-ph/0611385} {arXiv:astro-ph/0611385}
  \BibitemShut {NoStop}%
\bibitem [{\citenamefont {{M.V. Kossov}}(2002)}]{kossov}%
  \BibitemOpen
  \bibfield  {author} {\bibinfo {author} {\bibnamefont {{M.V. Kossov}}},\
  }\bibfield  {title} {\bibinfo {title} {Approximation of photonuclear
  interaction cross-sections},\ }\href
  {https://doi.org/10.1140/epja/i2002-10008-x} {\bibfield  {journal} {\bibinfo
  {journal} {Eur. Phys. J. A}\ }\textbf {\bibinfo {volume} {14}},\ \bibinfo
  {pages} {377} (\bibinfo {year} {2002})}\BibitemShut {NoStop}%
\bibitem [{\citenamefont {Globus}\ \emph {et~al.}(2015)\citenamefont {Globus},
  \citenamefont {Allard}, \citenamefont {Mochkovitch},\ and\ \citenamefont
  {Parizot}}]{Globus:2014fka}%
  \BibitemOpen
  \bibfield  {author} {\bibinfo {author} {\bibfnamefont {N.}~\bibnamefont
  {Globus}}, \bibinfo {author} {\bibfnamefont {D.}~\bibnamefont {Allard}},
  \bibinfo {author} {\bibfnamefont {R.}~\bibnamefont {Mochkovitch}},\ and\
  \bibinfo {author} {\bibfnamefont {E.}~\bibnamefont {Parizot}},\ }\bibfield
  {title} {\bibinfo {title} {{UHECR acceleration at GRB internal shocks}},\
  }\href {https://doi.org/10.1093/mnras/stv893} {\bibfield  {journal} {\bibinfo
   {journal} {Mon. Not. Roy. Astron. Soc.}\ }\textbf {\bibinfo {volume}
  {451}},\ \bibinfo {pages} {751} (\bibinfo {year} {2015})},\ \Eprint
  {https://arxiv.org/abs/1409.1271} {arXiv:1409.1271 [astro-ph.HE]}
  \BibitemShut {NoStop}%
\bibitem [{\citenamefont {Boncioli}\ \emph {et~al.}(2017)\citenamefont
  {Boncioli}, \citenamefont {Fedynitch},\ and\ \citenamefont
  {Winter}}]{Boncioli:2016lkt}%
  \BibitemOpen
  \bibfield  {author} {\bibinfo {author} {\bibfnamefont {D.}~\bibnamefont
  {Boncioli}}, \bibinfo {author} {\bibfnamefont {A.}~\bibnamefont
  {Fedynitch}},\ and\ \bibinfo {author} {\bibfnamefont {W.}~\bibnamefont
  {Winter}},\ }\bibfield  {title} {\bibinfo {title} {{Nuclear Physics Meets the
  Sources of the Ultra-High Energy Cosmic Rays}},\ }\href
  {https://doi.org/10.1038/s41598-017-05120-7} {\bibfield  {journal} {\bibinfo
  {journal} {Sci. Rep.}\ }\textbf {\bibinfo {volume} {7}},\ \bibinfo {pages}
  {4882} (\bibinfo {year} {2017})},\ \Eprint {https://arxiv.org/abs/1607.07989}
  {arXiv:1607.07989 [astro-ph.HE]} \BibitemShut {NoStop}%
\bibitem [{\citenamefont {Morejon}\ \emph {et~al.}(2019)\citenamefont
  {Morejon}, \citenamefont {Fedynitch}, \citenamefont {Boncioli}, \citenamefont
  {Biehl},\ and\ \citenamefont {Winter}}]{Morejon:2019pfu}%
  \BibitemOpen
  \bibfield  {author} {\bibinfo {author} {\bibfnamefont {L.}~\bibnamefont
  {Morejon}}, \bibinfo {author} {\bibfnamefont {A.}~\bibnamefont {Fedynitch}},
  \bibinfo {author} {\bibfnamefont {D.}~\bibnamefont {Boncioli}}, \bibinfo
  {author} {\bibfnamefont {D.}~\bibnamefont {Biehl}},\ and\ \bibinfo {author}
  {\bibfnamefont {W.}~\bibnamefont {Winter}},\ }\bibfield  {title} {\bibinfo
  {title} {{Improved photomeson model for interactions of cosmic ray nuclei}},\
  }\href {https://doi.org/10.1088/1475-7516/2019/11/007} {\bibfield  {journal}
  {\bibinfo  {journal} {JCAP}\ }\textbf {\bibinfo {volume} {11}},\ \bibinfo
  {pages} {007}},\ \Eprint {https://arxiv.org/abs/1904.07999} {arXiv:1904.07999
  [astro-ph.HE]} \BibitemShut {NoStop}%
\bibitem [{\citenamefont {L{\"u}}\ and\ \citenamefont
  {Zhang}(2014)}]{Lu:2014oda}%
  \BibitemOpen
  \bibfield  {author} {\bibinfo {author} {\bibfnamefont {H.-J.}\ \bibnamefont
  {L{\"u}}}\ and\ \bibinfo {author} {\bibfnamefont {B.}~\bibnamefont {Zhang}},\
  }\bibfield  {title} {\bibinfo {title} {{A Test of the Millisecond Magnetar
  Central Engine Model of Gamma-Ray Bursts with Swift Data}},\ }\href
  {https://doi.org/10.1088/0004-637X/785/1/74} {\bibfield  {journal} {\bibinfo
  {journal} {Astrophys. J.}\ }\textbf {\bibinfo {volume} {785}},\ \bibinfo
  {pages} {74} (\bibinfo {year} {2014})},\ \Eprint
  {https://arxiv.org/abs/1401.1562} {arXiv:1401.1562 [astro-ph.HE]}
  \BibitemShut {NoStop}%
\bibitem [{\citenamefont {David}(2015)}]{David:2015ura}%
  \BibitemOpen
  \bibfield  {author} {\bibinfo {author} {\bibfnamefont {J.~C.}\ \bibnamefont
  {David}},\ }\bibfield  {title} {\bibinfo {title} {{Spallation reactions: A
  successful interplay between modeling and applications}},\ }\href
  {https://doi.org/10.1140/epja/i2015-15068-1} {\bibfield  {journal} {\bibinfo
  {journal} {Eur. Phys. J. A}\ }\textbf {\bibinfo {volume} {51}},\ \bibinfo
  {pages} {68} (\bibinfo {year} {2015})},\ \Eprint
  {https://arxiv.org/abs/1505.03282} {arXiv:1505.03282 [nucl-ex]} \BibitemShut
  {NoStop}%
\end{thebibliography}%

\end{document}